\documentclass[a4paper,fleqn]{cas-dc}

\usepackage[numbers]{natbib}
\usepackage[ruled,linesnumbered]{algorithm2e}
\usepackage{booktabs}
\usepackage{url}
\usepackage{color}
\usepackage{stfloats}
\usepackage{float} 
\usepackage{subfigure}
\usepackage{tikz}
\newcommand*{\circled}[1]{\lower.7ex\hbox{\tikz\draw (0pt, 0pt)%
    circle (.5em) node {\makebox[1em][c]{\small #1}};}}

\def\tsc#1{\csdef{#1}{\textsc{\lowercase{#1}}\xspace}}
\tsc{WGM}
\tsc{QE}
\tsc{EP}
\tsc{PMS}
\tsc{BEC}
\tsc{DE}
\begin{document}
\begin{sloppypar}
\let\WriteBookmarks\relax
\def\floatpagepagefraction{1}
\def\textpagefraction{.001}
\shorttitle{HEGrid: High Efficient Multi-Channel Radio Astronomical Data Gridding Framework}
\shortauthors{Hao Wang et~al.}

\title [mode = title]{HEGrid: A High Efficient Multi-Channel Radio Astronomical Data Gridding Framework in Heterogeneous Computing Environments} 
% \tnotetext[1]{This material is based upon work supported by the Joint Research Fund in Astronomy (grant Nos. U1731125, U1731243) under a cooperative agreement between the National Natural Science Foundation of China (NSFC) and Chinese Academy of Sciences, NSFC grant No. 11903056, the Cultivation Project for FAST Scientific Payoff and Research Achievement of CAMS-CAS, as well as the Open Project Program of the Key Laboratory of FAST, NAOC, Chinese Academy of Sciences.}

\author[1]{Hao Wang}

\author[1]{Ce Yu}

\author[1]{Jian Xiao}
\ead{xiaojian@tju.edu.cn}
\cormark[1]

\author[1]{Shanjiang Tang}

\author[2]{Min Long}
\ead{minlong@boisestate.edu}
\cormark[1]
\cortext[cor1]{Corresponding author}

\author[3,4]{Ming Zhu}

\address[1]{College of Intelligence and computing, Tianjin University, No.135 Yaguan Rood, Haihe Education Park, Tianjin, 300350, China}
\address[2]{Department of Computer Science, Boise State University, Boise, ID 83725, USA}
\address[3]{National Astronomical Observatories, Chinese Academy of Sciences, 20A Datun Road, Chaoyang District, Beijing, 100101, China}
\address[4]{CAS Key Laboratory of FAST, National Astronomical Observatories, Chinese Academy of Sciences}

\begin{abstract}
The challenge to fully exploit the potential of existing and upcoming scientific instruments like large single-dish radio telescopes is to process the collected massive data effectively and efficiently. 
As a "quasi 2D stencil computation" with the "Moore neighborhood pattern," gridding is the most computationally intensive step in data reduction pipeline for radio astronomy studies, enabling astronomers to create correct sky images for further analysis.
However, the existing gridding frameworks can either only run on multi-core CPU architecture or do not support high-concurrency, multi-channel data gridding.
Their performance is then limited, and there are emerging needs for innovative gridding frameworks to process data from large single-dish radio telescopes like the Five-hundred-meter Aperture Spherical Telescope (FAST).
To address those challenges, we developed a \textbf{H}igh \textbf{E}fficient \textbf{Grid}ding framework, \textbf{HEGrid}, by overcoming the above limitations.  
HEGrid is the first effort to solve the gridding of multi-channel data from the large single-dish radio telescope by multi-pipeline concurrency in the CPU-GPU heterogeneous environment.
Specifically, we propose and construct the gridding pipeline in heterogeneous computing environments and achieve multi-pipeline concurrency for high performance multi-channel processing. 
Furthermore, 
%since computation and data correlation is centered in the proposed gridding framework, 
we propose pipeline-based co-optimization to alleviate the potential negative performance impact of possible intra- and inter-pipeline low computation and I/O utilization, including component share-based redundancy elimination, thread-level data reuse and overlapping I/O and computation.
Our experiments are based on both simulated datasets and actual FAST observational datasets. The results show that HEGrid outperforms other state-of-the-art gridding frameworks by up to 5.5x and has robust hardware portability, including AMD Radeon Instinct GPU and NVIDIA GPU.
\end{abstract}

% % Research highlights
% \begin{highlights}
% \item Research highlights item 1
% \item Research highlights item 2
% \item Research highlights item 3
% \end{highlights}


\maketitle

\section{Introduction}\label{section:introduction}
Effective and efficient data processing methods are an emerging need to fully exploit the potential of existing and upcoming scientific instruments and accelerate scientific discovery, such as data processing for the large single-dish radio telescopes FAST, Arecibo, Effelsberg and Green Bank, etc.
To record sky images from a wide range of frequencies, large single-dish radio telescope receivers contain a large number of independent channels with various band coverage settings.
Five-hundred-meter Aperture Spherical Telescope (FAST) \cite{bigot2015hi, dunning2017,li18,yue2012fast}, the world's largest single-dish radio telescope, has been in operation since 2020.
FAST receivers comprise 65,536 independent frequency channels (a significantly large but typical number for many large single-dish radio telescopes) and generate a massive volume of radio astronomical data across all frequency channels at a rate of 10-20 PB-size per year \cite{li18}. 

To obtain the correct sky images from such data, gridding is one of the critical steps which maps non-uniform data samples onto a uniformly distributed target grid map (referred to as \textbf{target map}) for further analysis. 
It is usually the most computationally intensive and time-consuming step \cite{griffin2018end, veenboer17, wang2020processing} due to the huge size of data in multiple channels. 
%the collected data in different frequency channels are independent to each other.
%这个数值不大，暂时不表示的好
%Existing work illustrated that with a 16-core CPU platform, it would take $\sim$ 21 CPU hours to perform gridding computation on 1 TB spectral line data using the CPU multi-thread method \cite{wang21}.  % \red(Should we say it? It means up to 420 k CPU-hrs, not too big? )  双盲审不能出现our previous之类的关键字，这里说的耗时长也不是说我们之前的工作的时间，而是在之前的文章中分析的CPU版本的运行时间，放在这里是为了体现数据量的庞大和gridding的耗时
For those reasons, there is a great need for fast and high-performance gridding frameworks to process multi-channel radio astronomical data from large single-dish radio telescopes.

Gridding algorithm is similar to stencil computation since it iteratively updates each target cell based on neighboring points and can be treated as one type of "quasi 2D stencil computation" with "Moore neighborhood pattern"\cite{traff2021mpi}. However, gridding also differs from the stencil computation in the following ways:
(1) the number of selected neighboring points for each cell may not be fixed but vary significantly;
(2) the number and location of neighboring points for each cell is not determined till the cell is updated.
These two features pose a challenge for effective access to neighboring points contributing to the calculation.
In addition, there would be much more neighboring points used in gridding than in the stencil computation.
For instance, in some gridding applications with high sampling densities, the number of neighboring points could reach nearly 90,000, adding additional challenge for cells update.

A number of gridding frameworks have been developed for processing data from various types of radio telescopes.
Among them, Cygrid \cite{winkel16} is one of the most popular and effective gridding frameworks. It supports multi-core CPU architecture and has been applied to the Effelsberg-Bonn HI Survey and the Galactic All-Sky Survey \cite{bekhti2016hi4pi}.
However, as discussed above, the gridding is more suitable for implementation on GPU architectures rather than CPU, due to its features of single instruction, multiple data stream (SIMD). 
Our previous work, HCGrid \cite{wang21}, gridding framework prototype designed in CPU-GPU heterogeneous computing environments for the large single-dish radio telescope, such as the FAST, which has demonstrated promising performance in the experiments with simulated datasets. 
However, HCGrid does not support high-concurrency processing of multi-channel data due to its low utilization of heterogeneous resources.
It is worth noting that there are other gridding algorithms used in CPU-GPU heterogeneous architectures, such as \cite{carcamo18}, \cite{merry16}, \cite{romein12}, \cite{veenboer17}, \cite{zhu20}, but none of them were designed for and can be applied to single-dish radio telescopes.
%However, all of them are suitable for radio telescope arrays but do not to single-dish radio telescopes like FAST.

%To achieve efficient multi-channel radio astronomy data gridding, designed for the large single-dish radio telescope, 

To overcome the limitations of existing gridding frameworks, combined with our previous work, we propose \textbf{HEGrid}, a high efficient gridding framework for the multi-channel data gridding of the large single-dish radio telescope. 
HEGrid is the first effort to solve the multi-channel data gridding of the large single-dish radio telescope by multi-pipeline concurrency in the CPU-GPU heterogeneous environment, it can port well for different GPU architectures including NVIDIA and AMD Radeon Instinct series.
Specifically, given gridding's computational correlation and data correlation, our contributions are:
\begin{enumerate}[(1)]
    \item We present the design of the HEGrid, including the building of the gridding pipeline and the multi-pipeline concurrency implementation.
    \item We propose pipeline-based co-optimization to alleviate the potential negative performance impact of possible low intra- and inter-pipeline computation and I/O utilization, which includes component share-based redundancy elimination, thread-level data reuse and overlapping I/O and computation.
    \item We port HEGrid to various GPU architectures, such as NVIDIA and AMD Radeon Instinct series, enabling HEGrid with robust hardware portability.
    \item We are releasing our implementation as open-source\footnote{https://github.com/HPCAstroAtTJU/HEGrid} for further research and use in achieving efficient gridding of astronomical data for current and upcoming large single-dish radio telescopes.
\end{enumerate}

The rest of the paper is organized as follows. 
We provide the background on gridding algorithms and the motivation for innovative methods in Section \ref{section2} and discuss the related work in Section \ref{section3}.
In Section \ref{section4}, we describe the design of HEGrid and optimization methods we used. 
Section \ref{experiments} compares HEGrid to other gridding frameworks by conducting various performance experiments using both simulated and actual observational data from FAST. 
Section \ref{section6} concludes the paper.

\section{Background and Motivation}\label{section2}
\subsection{The Need of Gridding}
Radio telescope consists of antenna and array receivers which detect radio signals from astronomical sources in the sky.
Given the size of the large single-dish radio telescope, deploying the telescope in a "drift scan" is usually needed as a feasible and near-optimum sky survey strategy. 
The "drift scan" means moving the telescope's receivers to a target azimuth and then fixing the telescope. 
Since the earth rotates once in 24 hours, various celestial objects enter the receiver's field of view once, recorded in a coordinate of two directions: right ascension and declination.

\begin{figure}
  \centering
  \includegraphics[width=\linewidth]{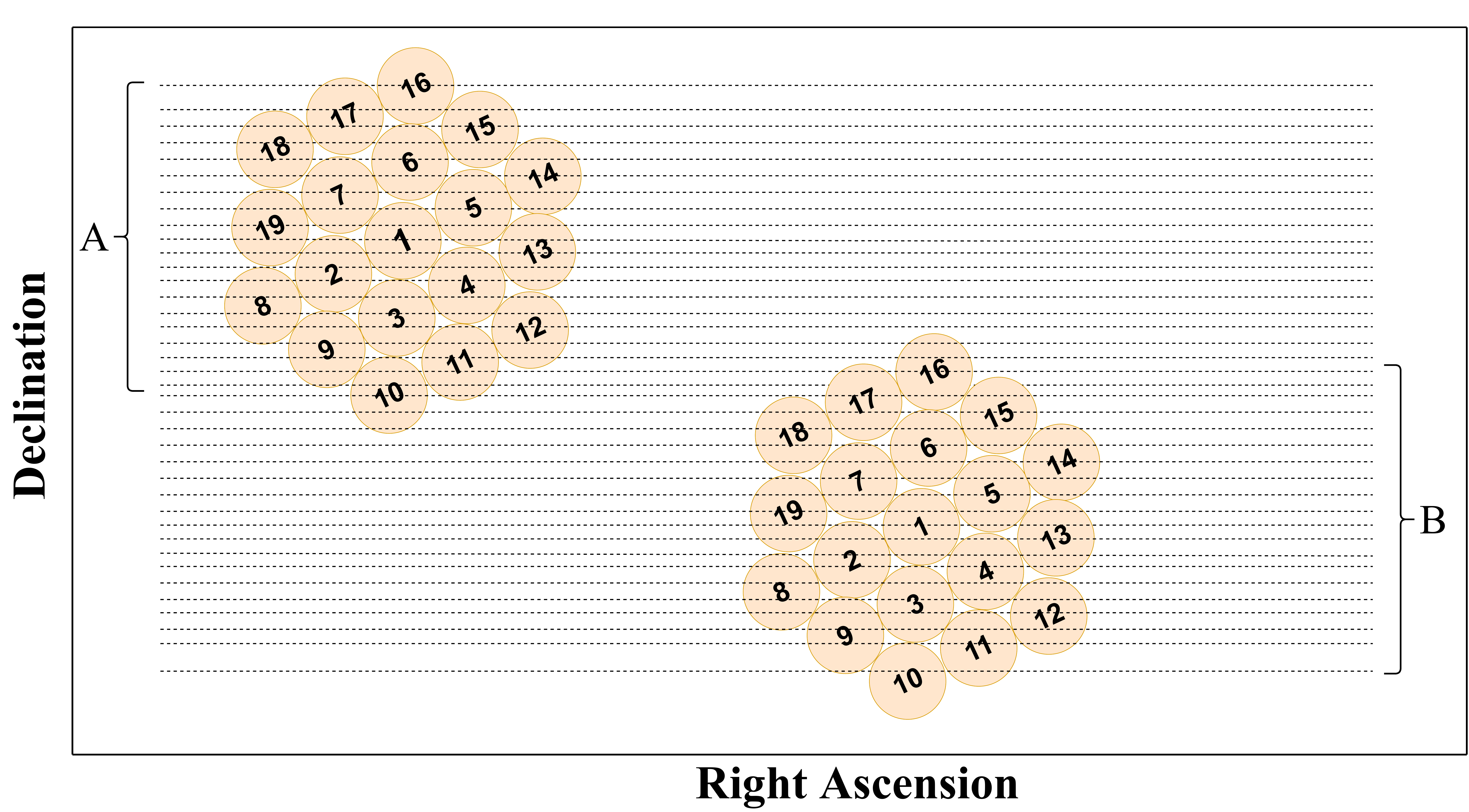}
  \caption{A schematic layout of 19-beam receiver of FAST and an example of two adjacent drift scans with areas A and B. The dotted lines show the drifting tracks of individual beams \cite{carrad2006,dunning2017, zhang2019status}.}
  \label{figure1}
\end{figure}

Figure \ref{figure1} shows the layout of 19-beam receivers of FAST and the "drift scan" strategy of FAST's survey \cite{dunning2017}.
FAST adopts a fixed rotation angle of the beam pattern toward a certain declination on the sky, and the drifting drags the receiver along the right ascension direction. By rotating the array by 23.4°, the most uniform coverage in the declination direction and super-Nyquist sampling can be achieved \cite{li18}.
After a 24-hour scan of a certain declination, a new declination is taken for new surveys of objects \cite{li18,zhang2019status}.
%----------------------------------------figure 1-------------------------%
%--------------------------------------------------------------------------%
%----------------------------------------figure 2-------------------------%
\begin{figure}
  \centering
  \includegraphics[width=\linewidth]{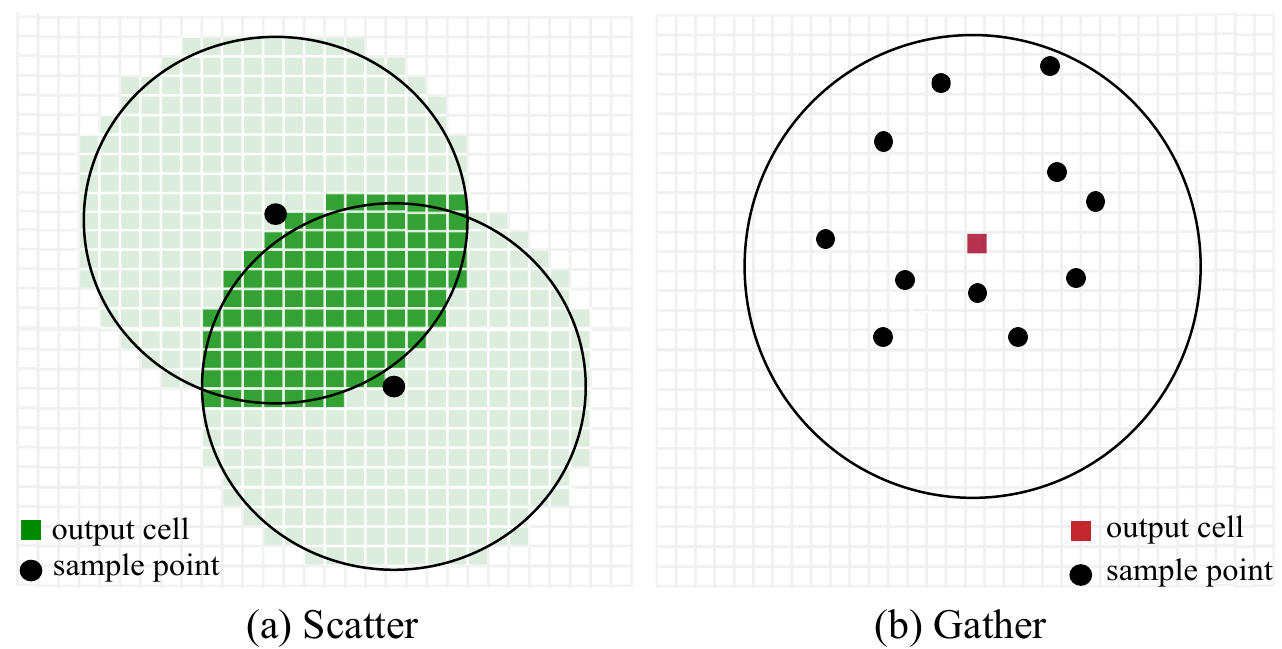}
  \caption{Input-oriented scatter (left) and output-oriented gather (right) gridding methods. Circles represent input data points and squares represent output cells.}
  \label{figure2}
\end{figure}
%--------------------------------------------------------------------------%
The received continuous data streams are stored in a multi-dimensional array according to values of channel number, right ascension, declination, forming a multi-dimensional datacube.
However, this strategy can cause a problem, the coverage of the raw recorded data has a much denser grid resolution in the right ascension direction than in the declination direction \cite{fabello2011alfalfa, giovanelli2005arecibo}, limiting itself to be directly used for scientific analysis, which requires a uniform interval in the two spatial directions of right ascension and declination to obtain high quality of sky image.
Like X-ray computed tomography \cite{blas2014surfing}, gridding can be seen as an image reconstruction method in radio astronomy, which solves the problem of uneven distribution of raw data points in the sampling space.

As with stencil computation, the kernel of a gridding computation typically contains a weighted sum of neighboring points for each target cell, and applies two methods of scatter or gather to the calculation \cite{zhao2019exploiting}:
\begin{enumerate}[(1)]
    \item Scatter: The scatter method traverses each input data point and broadcasts its sampling value to other output cells within the convolution kernel. Then the sampling value is weighted and updated to each target output cells, as shown in Figure \ref{figure2} (a).
    \item Gather: The gather method traverses each output cell and searches for neighboring points within its convolution kernel. Then it adds up weighted values and updates to the output cell, as shown in Figure \ref{figure2} (b).
\end{enumerate}
%----------------------------------------figure 3-------------------------%
\begin{figure*}
  \centering
  \includegraphics[scale=0.9]{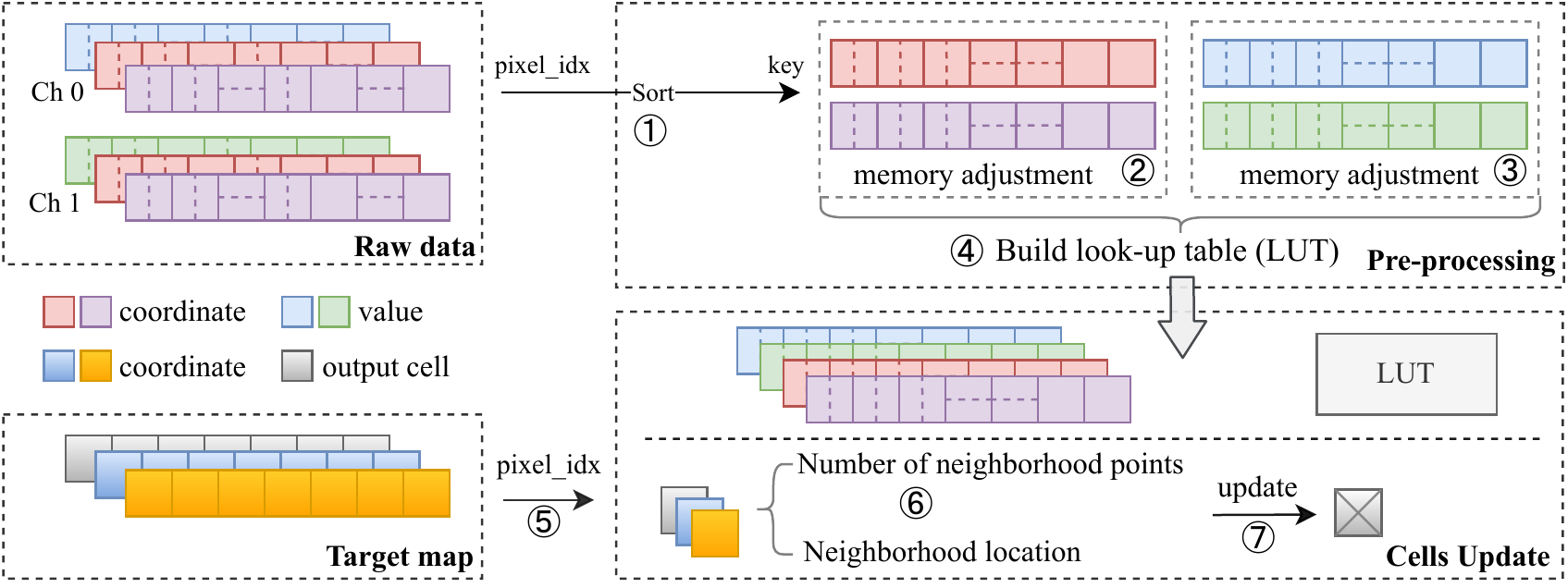}
  \caption{The overview of the HEGrid pipeline. The step of pre-processing runs on CPU. It first computes and sorts the pixel\_idx of the raw data points, then adjusts the raw data memory, building a look-up table. The step of cells update runs on GPU, which loads the target map, raw data, and LUT, then computes the contribution region and updates the target cell. ch0 and ch1 represent different frequency channels.}
  \label{figure3}
\end{figure*}
%--------------------------------------------------------------------------%

% The scatter method has the risk of “writing competition”, that is, multiple input data points might update the same output cell simultaneously and causes a conflict. 
% Although this issue can be prevented by using atomic or other synchronous operations \cite{mccool12, schweizer15, van09}, the degradation in parallelization could still be a problem.
% On the other hand, given upcoming GPU-based implementations, the gather method can take better advantage of GPU architectures, making itself a better choice for gridding than the scatter method. 

\subsection{The Gridding Algorithm}
The gather-based gridding algorithm is as follows. 
After determining a target grid map, the output value for each targeted grid cell is calculated as the weighted sum of all neighboring samples.
Let $ \mathbb{S}=\{s_1,s_2,\cdots,s_N\}$ denote $N$ discrete, non-uniformly spaced input samples in the right ascension-declination plane (ra-dec).
Each sample $s_n\in\mathbb{S},\ (n \in \{1,2,...,N\})$ has equatorial coordinates $(\alpha_n,\delta_n)$ (i.e., right ascension and declination) and a sampled value of $V[s_n]$.
For the output grid map, the ra-dec plane is divided into a regular, uniform grid with $I \times J$ cells as $\mathbb{G}=\{g_{1,1},g_{1,2},\cdots,g_{I,J}\}$.
For any cell $ g_{i,j}\in\mathbb{G} $ with central coordinates $(\alpha_{i,j}, \delta_{i,j})$, its re-sampled value $V[g_{i,j}]$ equals the weighted sum of raw data $\mathbb{S}$ related to $ g_{i,j} $.
     \begin{equation}
      V[g_{i,j}] = {\frac{1}{W_{i,j}}}
         \sum_{n}V[s_n]w(\alpha_{i,j},\delta_{i,j};\alpha_n,\delta_n).
         \label{convolution}
     \end{equation}
\noindent
$s_n \in \mathbb{S}$ represents any raw input sample with a weighted contribution to $ g_{i,j} $; 
$w(\alpha_{i,j},\delta_{i,j};\alpha_n,\delta_n)$ is a convolution kernel (weighting function) depending on positions of the output cell and raw data points, usually related to distances between input and output coordinates; and $W_{i,j}= \sum_{n}w(\alpha_{i,j},\delta_{i,j};\alpha_n,\delta_n)$ is the normalisation coefficient.

\section{Related Work}\label{section3}
Gridding is one of the most critical tasks in processing radio astronomical data such as pulsar data, spectral line data and so on.  
%To advance research in radio astronomy, 
Several gridding frameworks have been developed and customized for such data processing in the field of radio astronomy.
Cygrid \cite{winkel16} is one of the state-of-the-art works and runs only on the CPU platforms, which has been applied to studies like the Effelsberg-Bonn HI Survey and the Galactic All-Sky Survey\cite{bekhti2016hi4pi}.
However, its gridding performance is limited by the CPU architecture because it can't handle well the main features of gridding: single-instruction and multiple data (SIMD) stream. 
Modern parallel processors, such as GPU instead of CPU should be able to provide a better platform for such SIMD operations.

HCGrid \cite{wang21} is designed for gridding data from single-dish radio telescopes. It is based on CPU-GPU heterogeneous architecture and can achieve good performance for single-channel data. However, it does not support the high concurrency processing of multi-channel data due to its low utilization of heterogeneous resources.

Image-Domain Gridding \cite{veenboer17} implemented the gridding on both CPU and GPU, and has been deployed to the LOFAR (Low-Frequency Array) Central Processing center. It utilizes CUDA stream and related mathematics library to optimize the gridding. However, it cannot be applied to single-dish radio telescopes, and the portability of the algorithm is not desirable.

Other methods including \cite{carcamo18}, \cite{merry16}, \cite{romein12} and \cite{zhu20}. 
For instance, \cite{romein12} developed a work-distribution scheme for gridding using GPU, which can reduce the memory access time of computing device, and map observed samples onto a grid with high efficiency.
\cite{merry16} optimized the algorithm using thread coarsening strategy, which makes each thread to handle multiple samples simultaneously.
However, like Image-Domain Gridding, all those existing methods are only designed for radio telescope array but not fully applicable to the large single-dish radio telescopes.

\section{Design of HEGrid}\label{section4}
This section introduces the design and implementation of our multi-channel radio astronomical data gridding framework, HEGrid.
We first explain the design of the HEGrid pipeline and present the capabilities of multiple pipeline concurrency on heterogeneous architecture. 
Then, the details of pipeline-based co-optimization strategies were given. 
Furthermore, we port HEGrid to heterogeneous computing environments with different GPU architectures.

\subsection{HEGrid Pipeline}\label{section:pipeline}
Figure \ref{figure3} shows the pipeline of HEGrid.
First, an efficient look-up table (LUT) is built in the pre-processing step to accelerate the contribution point acquisition process. 
Second, to accelerate the most time-consuming cell update step, we achieve cell update parallelization by using SIMT instruction-level parallelism on GPU. 

\subsubsection{Building the Efficient LUT}\label{section:LUT}
\begin{figure}
  \centering
  \includegraphics[width=6.5cm]{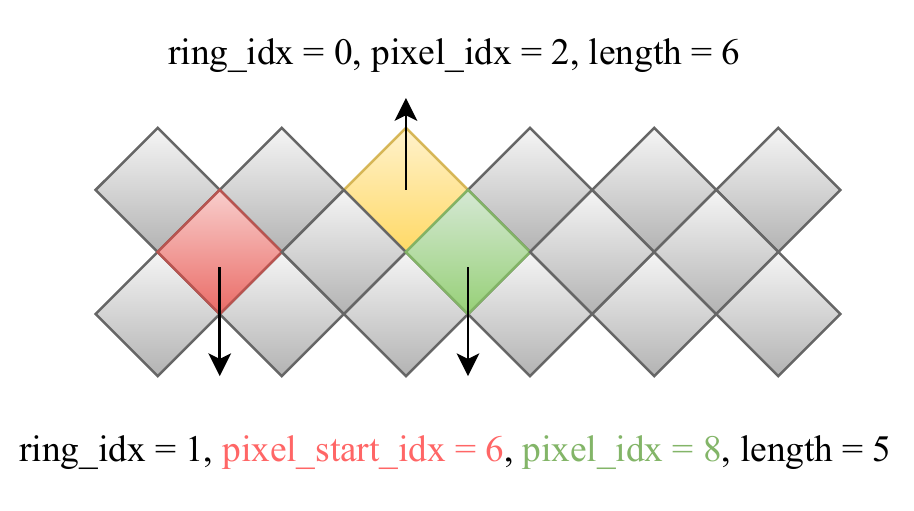}
  \caption{Raw data points are partitioned into different pixels. Each pixel has its index information, including pixel\_idx, ring\_idx, ring length, etc. }
  \label{figure4}
\end{figure}

As discussed in Section \ref{section:introduction}, uncertainty (location, number) of contribution points brings challenges to the cell update. 
We design an efficient lookup table with the help of HEALPix\footnote{HEALPix is a software package for hierarchical equal-area isolatitude pixelation on spherical surfaces in astronomy, which makes fast, accurate statistical or astrophysical analysis of massive all-sky datasets.}\cite{gorski05}.
With HEALPix, the raw data points on the celestial surface are partitioned into different pixels with different indexes, as shown in Figure \ref{figure4}.
%----------------------------------------figure 4-------------------------%
%--------------------------------------------------------------------------%

\begin{figure}
  \centering
  \includegraphics[width=\linewidth]{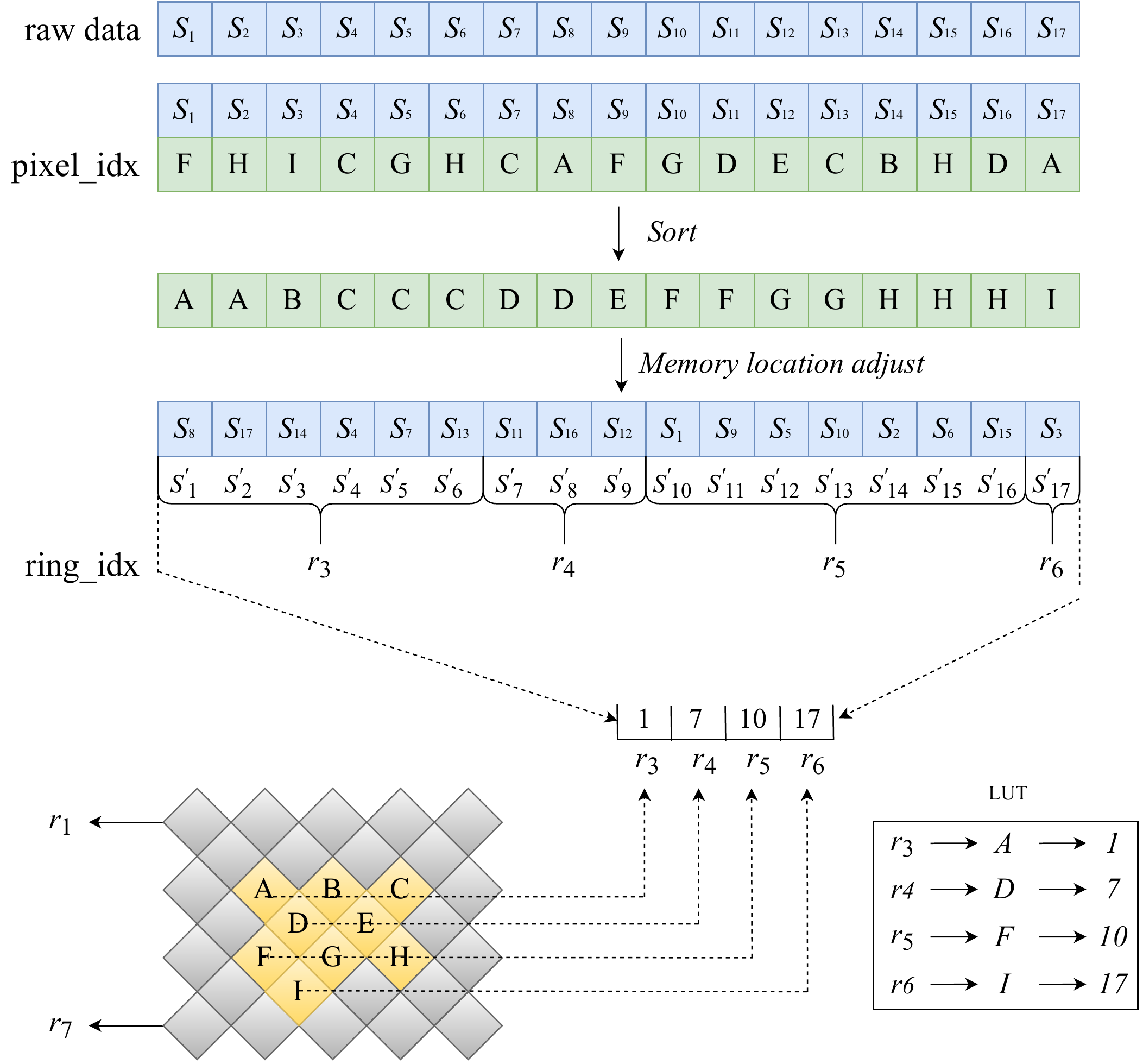}
  \caption{The schematic diagram of pre-processing. Top: Partition 17 raw data points $S_i(i=1,2...,17)$ into 9 pixels from "A" to "I". Middle: Adjust memory location of the raw data points based on the sorted pixel\_idx. Bottom: Build a lookup table based on the partitioned pixels.}
  \label{figure5}
\end{figure}

Figure \ref{figure5} gives an example for illustrating the process of lookup table build in Figure \ref{figure3}.
First, the 17 raw data points $S_i (i=1,2...,17)$ are partitioned into 9 pixels (A $\sim$ I), and then the pixel\_idx was sorted as shown in Figure \ref{figure5} (step \circled{1} in Figure \ref{figure3}). 
The Block Indirect sort algorithm is utilized in our work, its average time complexity is $\mathcal{O}(N log N)$.
Second, the location of coordinates and sampling value in memory for the raw data points was adjusted according to their pixel\_idx (steps \circled{2}, \circled{3}).
Third, after computing the ring\_idx of the latitude ring where different pixels are located, the lookup table is built based on the ring\_idx, pixel\_idx, and sampling points index (step \circled{4}).
The pre-processing step runs on CPU because there is a series of logic operations.
%----------------------------------------figure 5-------------------------%
%--------------------------------------------------------------------------%

\subsubsection{Parallelizing Cell Updates}\label{section:cell update}
Vectorization is a common technique employed in parallel processors.
Cell update step has computational characteristics of single instruction multiple data streams. In HEGrid, we manually vectorize the cell update on GPU in a SIMT manner.

%-----------------------------------Algorithm 1----------------------------------%
\begin{algorithm}
\caption{Cell Update Workflow}\label{algorithm1}
\KwIn{sorted data, LUT, target map}
\KwResult{updated cells}
\For{$target\_cell[0]$ \KwTo $target\_cell[n]$}{
Compute the pixel\_idx of the target cell\;
Compute the min contribution ring $ring\_{min}$\;
Compute the max contribution ring $ring\_{max}$\;
\For{$ring\_{min}$ \KwTo $ring\_{max}$}{
Compute the min contribution pixel $pixel\_{min}$\;
Compute the max contribution pixel $pixel\_{max}$\;
Compute the min indices $i$ of raw data in $pixel\_{min}$\;
Load the contribution points raw\_data[]\;
\While{pixel\_idx of $raw\_data[i]\leq pixel\_{max}$}{
\If{$d(target\_cell[],raw\_data[i]) \leq R$}{
Compute the weight sum\;
Compute the weighted value\;}
$i = i + 1$}
}
Normalize the weighted value\;
Update cell\;}
\end{algorithm}
%--------------------------------------------------------------------------------%
Algorithm \ref{algorithm1} shows the workflow of the cell update step.
After loading the data points and lookup table from the host, we first compute the pixel\_idx of the target cell, and determine the region (including the range of the ring, the starting pixel index on the contribution ring, and the offset between different contribution rings) of the contribution points for the target cell.
Then, with the help of the developed lookup table in Section \ref{section:LUT}, the contribution points are loaded ring-by-ring from device memory to the streaming multiprocessor (SM)\footnote{We mainly use terminology from NVIDIA hardware, such as SM, block, and warp, equivalent to CU, workgroup, and wavefront in AMD terminology.}, and their weights contributed to the target cell is computed.
When a thread finishes its task for one target cell, we cache the temporary results to register memory.

\begin{figure}
  \centering
  \includegraphics[width=\linewidth]{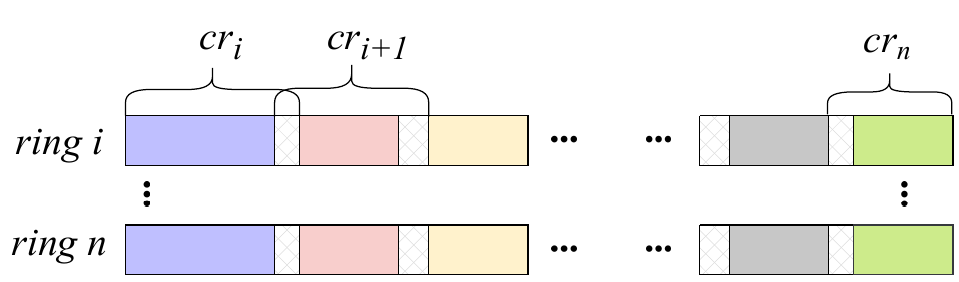}
  \caption{The location in memory of the contribution points for adjacent target cells on different rings. Different colors rectangles represent the contribution regions of diﬀerent target cells, and shaded rectangles are the contribution points shared by the adjacent target cells.}
  \label{figure6}
\end{figure}

\begin{figure}
  \centering
  \includegraphics[width=\linewidth]{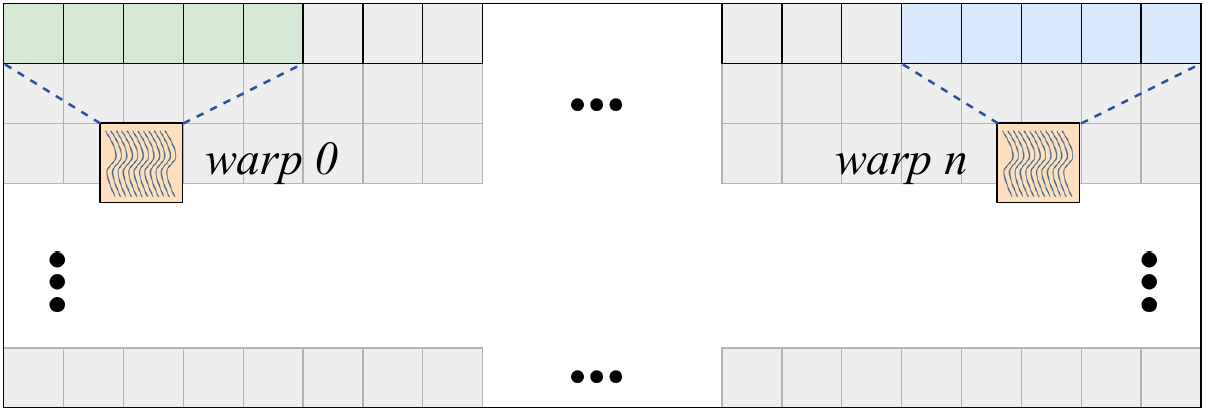}
  \caption{HEGrid parallel threads assignment in GPU. $n\times warp$ will be responsible for one-row target cells.}
  \label{figure7}
\end{figure}

As the smallest unit of SM execution and GPU resource scheduling for NVIDIA GPU and AMD GPU, the thread warp is the key to achieve efficient cell update.
In a warp, all threads execute in a single-instruction, multiple-thread (SIMT) manner \cite{jung2021snurhac}.
Target cells on the same row have the same contribution ring and the difference is that its contribution points may locate in different regions of the contribution ring.
Furtherly, as shown in Figure \ref{figure6}, the contribution points on the same contribution ring for adjacent target cells have overlapping contribution regions. 
To enable HEGrid could port to the computing environments with different GPU architectures and obtain a high cache hit rate on GPU, we propose an efficient organization strategy for parallel threads on GPU.
The detail is that we use one thread block as a vector, and each of the threads in the thread block is responsible for one target cell.
In addition, we assign the parallel threads along the longitude direction, and each thread warp is responsible for the computational tasks of the consecutive target cells.
Figure \ref{figure7} shows the parallel threads assignment in HEGrid.
The warp number in each row equals cell\_num\_one\_row / 32 (64)\footnote{The warp size (wavefront) in AMD GPU is 64.}.
By organizing parallel threads in this way, using warp as the task parallelization unit, we can strengthen the portability of the HEGrid across different GPU architectures.
Furthermore, our thread assignment strategy also considers the data reusability of inter-threads. 
The threads responsible for adjacent target cells can reuse the data cached in the GPU L1/L2 cache.
%----------------------------------------figure 6-------------------------%
%--------------------------------------------------------------------------%
%----------------------------------------figure 7-------------------------%
%--------------------------------------------------------------------------%

\subsection{Multi Pipeline Concurrency}
As discussed in the Section \ref{section:introduction}, because receivers typically cover a wide range of frequencies, the sky survey of large single-dish radio telescopes collect sky data using a large number of, independent frequency channels.
Thus, the data processing in those channels are naturally independent to each other. 
Combining multi-channel radio astronomical data characteristics, we explore the process-level parallelization and implementation of multi-channel gridding in this section.

\subsubsection{Profiling}
GPU supports multi-stream parallel execution \cite{durrani2021accelerating,jain2020crac, wang2021exploring}, which can facilitate the HEGrid to realize process-level parallelization on GPU, by dispatching the cell update for different channels to different streams.
%We analyze the timeline of HEGrid by experiment.

\begin{figure}
  \centering
  \includegraphics[width=\linewidth]{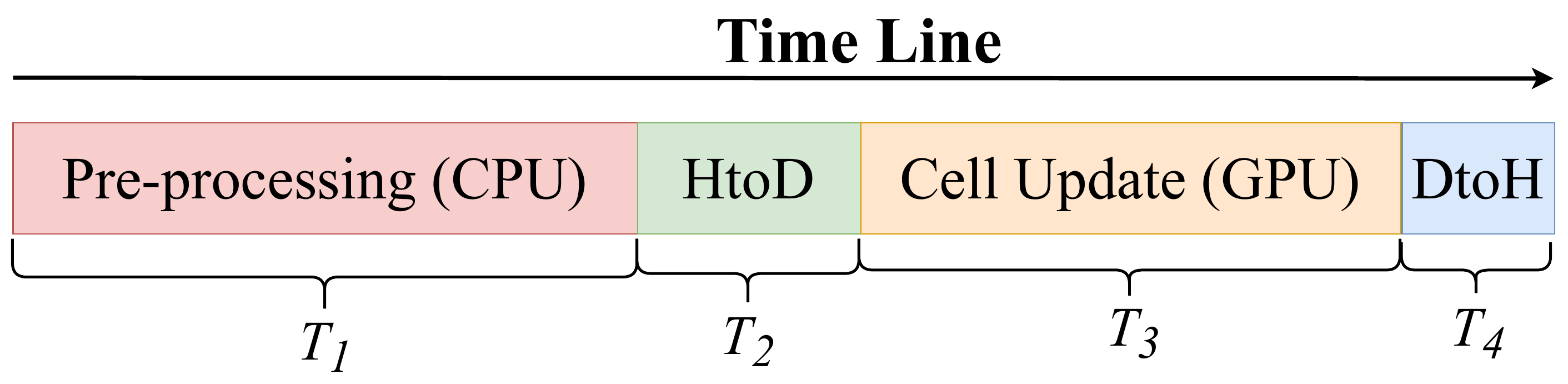}
  \caption{The experimental timeline of the HEGrid pipeline. HtoD and DtoH: "Host to Device" and "Device to Host". }
  \label{figure8}
\end{figure}

We analyzed the time spent at each stage of the HEGrid pipeline. 
Figure \ref{figure8} shows our experimental results. The length of the rectangle represents the duration. It can be seen that $T1 \ $\textgreater$ \ T3 \ $\textgreater$ \ T2 \ $\textgreater$ \ T4$.
%----------------------------------------figure 8-------------------------%
%--------------------------------------------------------------------------%

The prerequisite of using GPU streams in the gridding computation is that each stream executed concurrently should have access to data processed from the CPU, i.e., the CPU has to provide sufficient channels of data to each stream.
Therefore, while the CPU is still processing data from multiple channels using sequential execution, multiple GPU streams can execute concurrently only if $T1 \  + \ T2 \  $\textless$ \ T3$. 
Otherwise there will be idle streams waiting for data and degenerated to serial execution. 
However, the timeline in the HEGrid ($T1 \  + \ T2 \  $\textgreater$ \ T3$) is exactly opposite to the prerequisites (i.e., $T1 \  + \ T2 \  $\textless$ \ T3$), which prevented performance improvement using GPU streams. 

\subsubsection{Pipeline Concurrency and Scheduling}
\begin{figure*}
  \centering
  \includegraphics[width=\hsize]{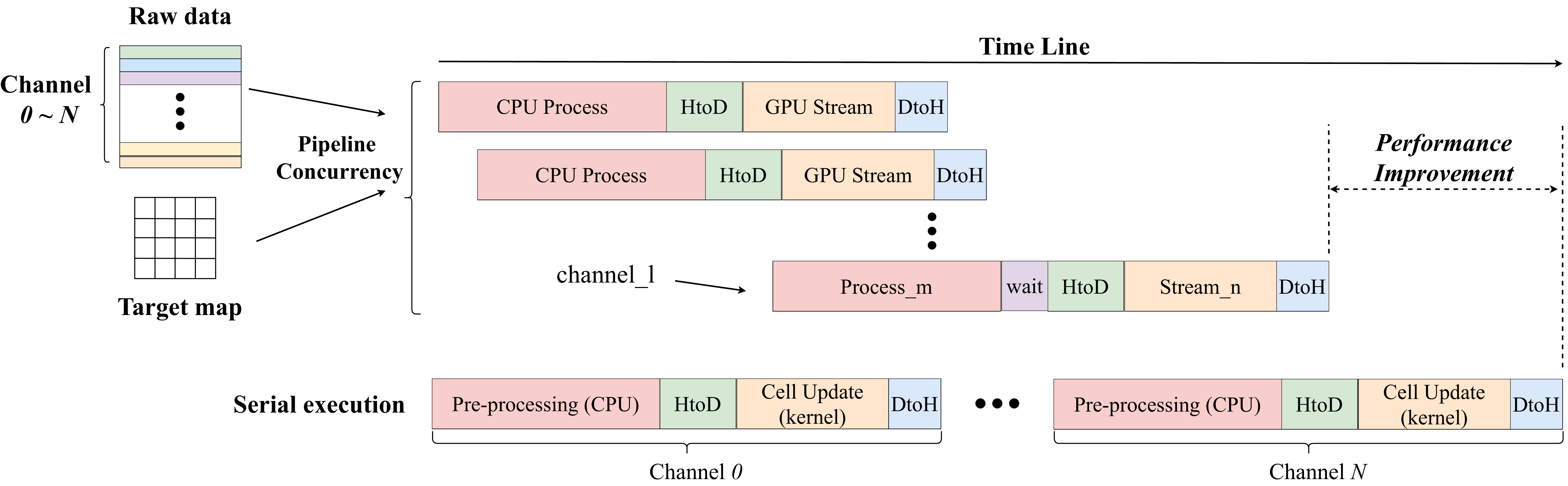}
  \caption{Multi pipeline concurrency on heterogeneous architecture. "wait" or idle stream waiting for data is mainly determined by the GPU's hardware transfer mechanism, where when two adjacent streams in the same direction (e.g., from the host to the device) are requesting to transfer data, the stream which requests first could block the other stream.}
  \label{figure9}
\end{figure*}

As we analyzed, partial parallelization of the HEGrid pipeline can only achieve in some instances. Otherwise, it will degrade to serialize.
In short, if there is concurrency in the pre-processing and data transfer, different GPU kernels can start asynchronously.
To achieves process-level parallelization of gridding, we propose multi pipeline concurrency on heterogeneous architectures, shown in Figure \ref{figure9}.
%----------------------------------------figure 9-------------------------%
%--------------------------------------------------------------------------%

By combining CPU multi-process and GPU multi-stream, we achieve multi pipeline concurrency.
The pipeline scheduling needs to be considered at both inter-and intra-pipeline levels. 
We found through experimental analysis that the data processing time at each stage, such as pre-processing, cell update, and the overall, is similar for different channels.
Therefore, the optimal two-level scheduling policy followed by pipeline should be FIFO.
As shown in Figure \ref{figure9}, the data from channel\_l will load to the idle process\_m, and the idle stream\_n will process the data from process\_m.
% Compared to serial execution, multi-pipeline concurrency is bound to yield performance improvements. 

\subsection{Pipeline-based Co-optimization}
\subsubsection{Component Share-based Redundancy Elimination}
\begin{figure}
  \centering
  \includegraphics[width=\linewidth]{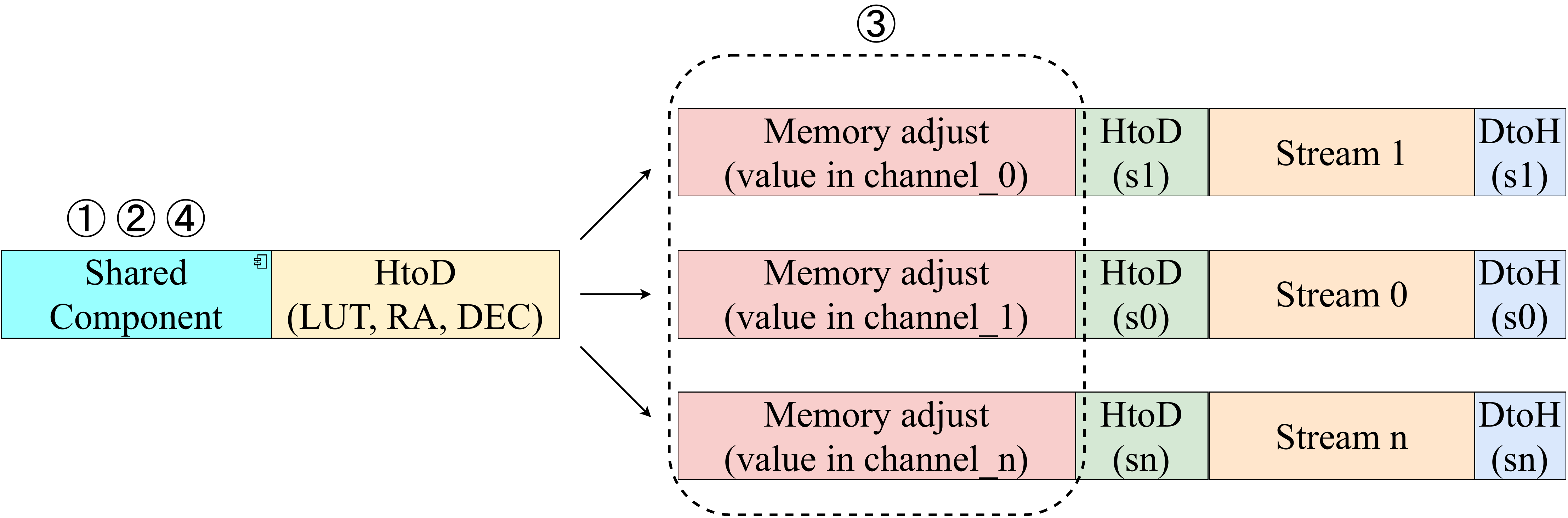}
  \caption{Component share-based redundancy elimination. The shared component is responsible for pixel\_idx computation and sort, adjustments of coordinates storage in memory and lookup table construction.}
  \label{figure10}
\end{figure}

As mentioned in Section \ref{section:pipeline}, data points in different channels with the same coordinate correspond to the same HEALPix pixels.
In HEGrid, each pipeline can build up its lookup table and load it from host to device.
It's not hard to see that this causes the existence of redundant computations and redundant data transfers.
We design a "shared component" mechanism to eliminate the duplicate construction of the lookup table.
The pre-processing steps plotted in Figure \ref{figure3} are divided into stages as shown in Figure \ref{figure10}. 
Steps \textbf{\circled{1}, \circled{2}, \circled{4} } are assigned to shared components because they can be reused in all pipelines.
Furthermore, a fixed size memory was allocated in the device, and the LUT, coordinates, and sampling value were loaded only once from the host to the device.
In the concurrency pipeline, the data from the shared component will be broadcasted to the cell update kernel of each pipeline.
%----------------------------------------figure 10-------------------------%
%--------------------------------------------------------------------------%

\subsubsection{Asynchronous Data Transfer and Computation}
As the number of concurrency pipelines increases, the number of data exchanges between CPU and GPU will also increase accordingly.
In order to reduce the cost of memory allocation, we implement a memory pool that can be reused by the gridding kernel. 
Each GPU stream can access its fraction of the memory pool through its stream id. 
We then allocate pinned memory on the host side to obtain the highest bandwidth.
Since the host CPU feeds cell update tasks to GPU asynchronously, the computation and data transfer can be overlapped.

\subsubsection{Thread-level Data Reuse}
In some gridding applications with high sampling densities, the number of neighboring points for one target cell can achieve 90,000, which challenges data transfer between the SM and device memory.
From experiment analysis, adjacent cells on the same row have the similar contribution region at high output resolution.
To increase the data reusability between different threads, we assign threads to be responsible for the computational tasks of multiple adjacent cells.  
In addition, the adjacent cells corresponding to the same thread will use the same intermediate results in the computation, such as the contributing rings and starting contributing points, etc.

\subsection{Porting to different GPU architectures}\label{section:porting}
With the advent of numerous accelerators with different architectures, hardware portability is becoming a feature that should be present in typical scientific applications.
As a data-driven astronomical application, gridding should have portability on different GPU architectures.
ROCm\footnote{https://rocmdocs.amd.com/en/latest/index.html}\footnote{https://github.com/ROCm-Developer-Tools} is AMD’s open-source software platform for GPU-accelerated high-performance computing and machine learning \cite{abdelfattah2021gpu, leinhauser2022metrics, otterness2020amd}.
An efficient thread assignment scheme is designed for the cell update step on GPU, in Section \ref{section:cell update}, to enable HEGrid could adapt to AMD GPU and obtain high performance.
Through the ROCm platform, we successfully port HEGrid to the AMD GPU architecture and run it on the Instinct MI50 GPU.
In addition, HEGrid can also run on the GPU with AMD's latest CDNA architecture.
There are challenges in porting HEGrid to the AMD GPU.
First, ROCm does not support texture memory bound, we allocate global memory instead. 
Second, the performance profiling tools integrated with ROCm are under development and are not yet perfect. 
Currently, we manually tune the size of thread blocks to get better performance and have not yet done further profiling and optimization of performance based on architecture, which is part of our future work.

\section{EXPERIMENTS}\label{experiments}
In this section, we perform detailed evaluations of HEGrid, in aspects of overall performance, the performance impact of the optimization scheme, the portability on different GPU architectures, and the accuracy of HEGrid.  
We used both simulated datasets and actual observational datasets from FAST in the experiment and compared the HEGrid to other gridding frameworks.  

%-----------------------------------Table 1----------------------------------%
\begin{table*}[width=\textwidth,cols=4,pos=h]
  \caption{Hardware configurations for experiments.}
    \begin{tabular}{lrccrcc}
    \toprule
    Server  &       & \multicolumn{2}{c}{Server\_V} &       & \multicolumn{2}{c}{Server\_M} \\
\cmidrule{3-4}\cmidrule{6-7}    Processor &       & CPU   & GPU   &       & CPU   & GPU \\
    \midrule
    Model &       & Xeon Gold 6151 & Tesla V100 &       & Xeon  E5-2620 & Instinct MI50 \\
    Transistors &       & 14nm  & 12nm  &       & 22nm  & 7nm \\
    \# Cores &       & 16    & 5120  &       & 32    & 3840 \\
    Base Frequency (MHz) &       & 3000  & 1245  &       & 2100  & 1200 \\
    Max Frequency (MHz) &       & 3400  & 1380  &       & 3000  & 1746 \\
    Device Memory Type &       & -     & HBM2  &       & -     & HBM2 \\
    Device Memory Clock (MHZ) &       & -     & 876   &       & -     & 1000 \\
    Device Memory Bandwidth (GB/s) &       & -     & 897   &       & -     & 1024 \\
    Device Memory Size (GB) &       & -     & 16    &       & -     & 16 \\
    Host Memory Size (GB) &       & 128   & -     &       & 128   & - \\
    \bottomrule
    \end{tabular}%
  \label{table1}%
\end{table*}%

%--------------------------------------------------------------------------------%

\subsection{Experimental Setup}
\subsubsection{Hardware Configurations}
Experiments are conducted on two servers with different GPU architectures, namely Server\_V with Xeon Gold 6151 CPU and NVIDIA V100 GPU, and Server\_M with Xeon E5-2620 CPU and AMD MI50 GPU. Their hardware configurations are shown in Table \ref{table1}.
%-----------------------------------Table 2----------------------------------%
\begin{table}
  \centering
  \caption{Datasets for Experiment.}
  \resizebox{\linewidth}{!}{
    \begin{tabular}{lcc}
    \toprule
    Dataset  & Simulated & Observed (by FAST) \\
    \midrule
    File Format & HDF5  & HDF5 \\
    Beam Size & $180^{\prime \prime}$   & $180^{\prime \prime}$ \\
    Map Size & $60^\circ \times 20^\circ$ & $60^\circ \times 20^\circ$ \\
    Map Center & $(30^\circ, 41^\circ)$ & $(30^\circ, 41^\circ)$ \\
    Points Num & $1.50E+07 \sim 1.90E+07$ & 2.83E+06 \\
    Channels Num & 50    & $10 \sim 50$ \\
    File Size & $2.91 \sim 3.68$ GB & 3.32 GB \\
    \bottomrule
    \end{tabular}}%
  \label{table2}%
\end{table}%

%--------------------------------------------------------------------------------%
\subsubsection{Datasets}
As performance could be data dependent, we use two different datasets for the performance evaluation, as shown in Table \ref{table2}. 
The first is a simulated datasets generated with the observation parameters of FAST. 
The second is an actual observational datasets collected by FAST. 
It is worth noting that the simulated dataset differs from the actual dataset with a much larger data size at each channel ($10^7$ vs $10^6$).
This is because FAST has not yet completed a full survey of the sky, and it requires multiple repeat scans of the same target sky area to get as complete a picture of the sky as possible. 
Additionally, FAST will complete a more comprehensive survey in the coming period, and more complete data are not yet available for performance analysis, so we use simulated data with high sampling density (i.e., the datasets with large data sizes) for the performance analysis in some experiments.

\subsubsection{Performance Metrics}
We use speedup as our main metric to measure the performance changes.
In Section \ref{section:overall} and \ref{section:5.4}, the speedup is the ratio of the baseline running time to the running time of HEGrid. 
A state-of-the-art gridding framework with the shortest running time (i.e., the best) would be selected as the baseline.
In Section \ref{section:5.3}, the speedup is the ratio of the running time of the non-optimized HEGrid (i.e., baseline) to the running time of the optimized HEGrid.

\subsection{Overall Performance}\label{section:overall}
The comparison between HEGrid and other state-of-the-art gridding frameworks is shown in Table \ref{table3}.
It can be seen that HEGrid outperforms other frameworks by up to 5.5x performance speedup in all experiments. This demonstrated the advantage of HEGrid in radio astronomy data gridding.
%-----------------------------------Table 3----------------------------------%

\begin{table*}[width=\textwidth,cols=4,pos=h]
  \centering
  \caption{Comparison of the performance of gridding frameworks (Running Time (s)).}
  \resizebox{\textwidth}{!}{
    \begin{tabular}{lccccccccccc}
    \toprule
    Dataset & \multicolumn{5}{c}{Simulated}         &       & \multicolumn{5}{c}{Observed (by FAST)} \\
\cmidrule{2-6}\cmidrule{8-12}    Datasize / Channel num & 1.50E+07 & 1.60E+07 & 1.70E+07 & 1.80E+07 & 1.90E+07 &       & 10    & 20    & 30    & 40    & 50 \\
    \midrule
    Cygrid & 165.87 & 171.37 & 178.99 & 187.31 & 194.6 &       & 77.77 & 79.01 & 80.12 & 80.97 & 84.76 \\
    HCGrid & 173.25 & 178.82 & 189.01 & 196.43 & 206.28 &       & 25.5  & 52.02 & 79.58 & 113.12 & 137.1 \\
    HEGrid & 30.21 & 32.77 & 35.27 & 38.2  & 40.94 &       & 7.15  & 12.77 & 18.29 & 24    & 29.6 \\
    \midrule
    Speedup (HEGrid) & \textbf{5.49} & \textbf{5.23} & \textbf{5.07} & \textbf{4.90} & \textbf{4.75} &       & \textbf{3.57} & \textbf{4.07} & \textbf{4.35} & \textbf{3.37} & \textbf{2.86} \\
    \bottomrule
    \end{tabular}}%
  \label{table3}%
\end{table*}%
%---------------------------------------------------------------------------%

Specifically, on the simulated dataset, we evaluate the dependence of performance on data size per channel. HEGrid has the best performance compared to Cygrid and HCGrid. It's 5.5x faster than Cygrid. This benefits from our design of HEGrid running on CPU-GPU heterogeneous architectures and the high process-level parallelism in the gridding. 

On the FAST's observed data, we evaluate the dependence of performance on the number of frequency channels. HEGrid also outperforms Cygrid and HCGrid.  It's 4.3x faster than HCGrid. This demonstrates the effectiveness of our parallelization strategy of multi pipeline concurrency and other optimization techniques.

\subsection{Analysis of Performance Optimizations}\label{section:5.3}

\subsubsection{Redundancy Elimination}
\begin{figure}
  \centering
  \includegraphics[width=\hsize]{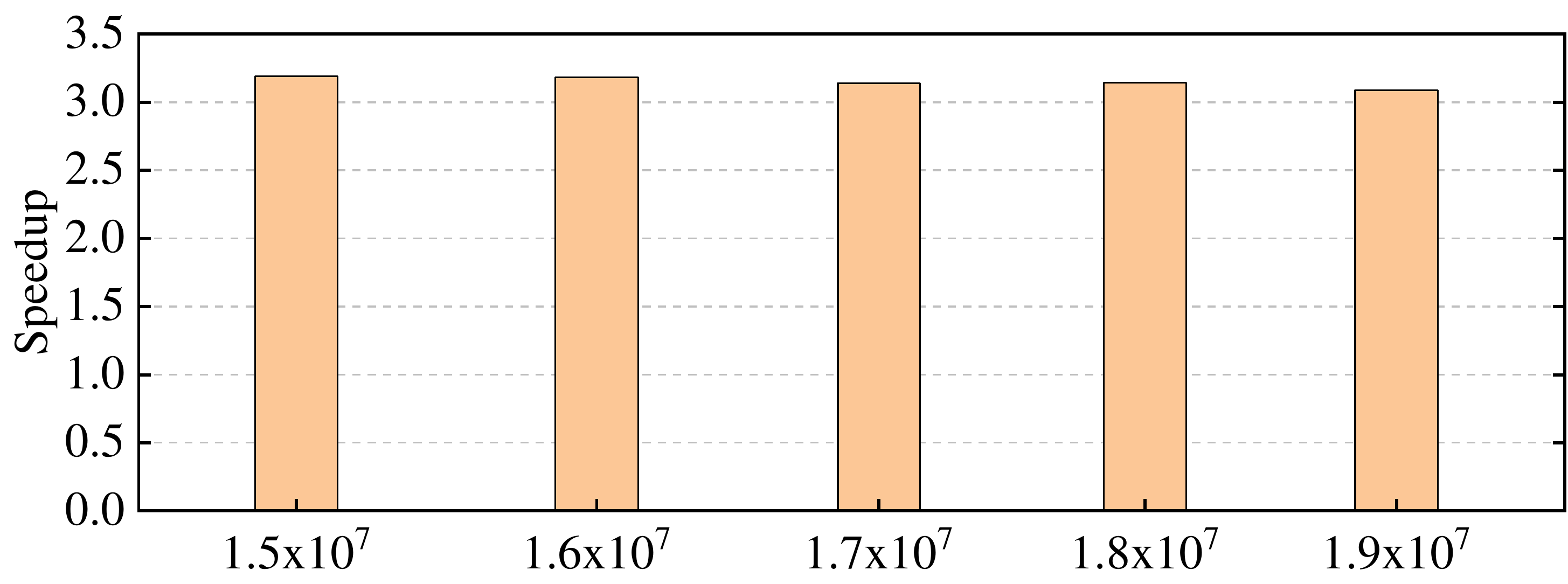}
  \caption{Performance improvement of redundancy elimination scheme under the simulated datasets with different data sizes.}
  \label{figure11}
\end{figure}

\begin{figure}
  \centering
  \includegraphics[width=\hsize]{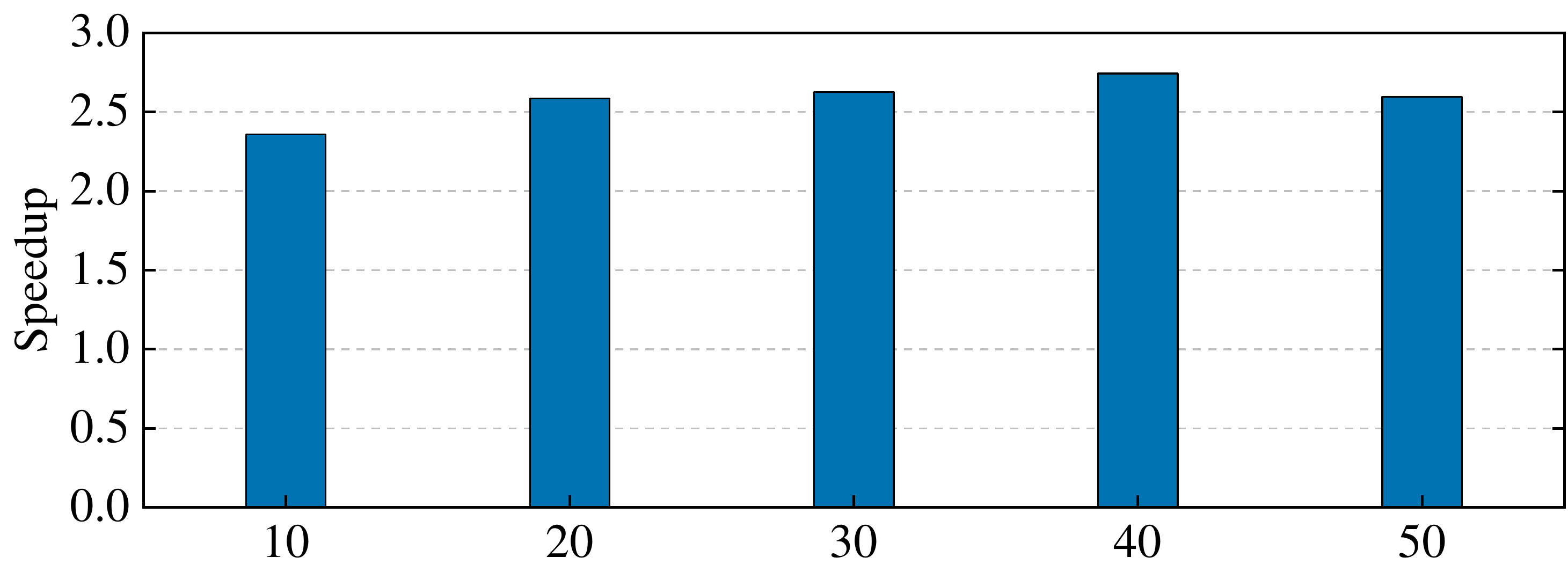}
  \caption{Performance improvements of redundancy elimination scheme under the FAST's observed data with different frequency channels.}
  \label{figure12}
\end{figure}

Our optimization scheme on the CPU eliminates the duplicate computations in the pre-processing step.
We now measure the performance of HEGrid under the optimization of component share-based redundancy elimination.
Figure \ref{figure11} and Figure \ref{figure12} show the overall performance improvements brought by our scheme on the simulated dataset and FAST's observed data respectively.

First, under the simulated datasets with different data sizes, the average performance improvement brought by the redundancy elimination is 3.2x.
Specifically, compared with Figure \ref{figure12}, performance improvement of redundancy elimination scheme is more obvious for large datasets. 
This is because the duplicate lookup table construction and the duplicate data loading from the host to the device will be one of the major performance challenges in processing the large datasets.
Therefore, the performance benefits of the redundancy elimination strategy will be more obvious.
Second, the performance improvement is also evident in the data observed by FAST with different frequency channels, demonstrating the effectiveness of our strategy for multiple pipeline concurrency.
In addition, the performance improvement exhibited with a channel count of 50 is slightly lower than that of Figure \ref{figure11}, which further proves the performance advantage of the redundancy elimination strategy for large datasets, the scale of data from FAST will need to handle in the future.
%----------------------------------------figure 11-------------------------%
%--------------------------------------------------------------------------%
%----------------------------------------figure 12-------------------------%
%--------------------------------------------------------------------------%

\subsubsection{Threads Concurrency on GPU}
\begin{figure}
  \centering
  \includegraphics[width=\linewidth]{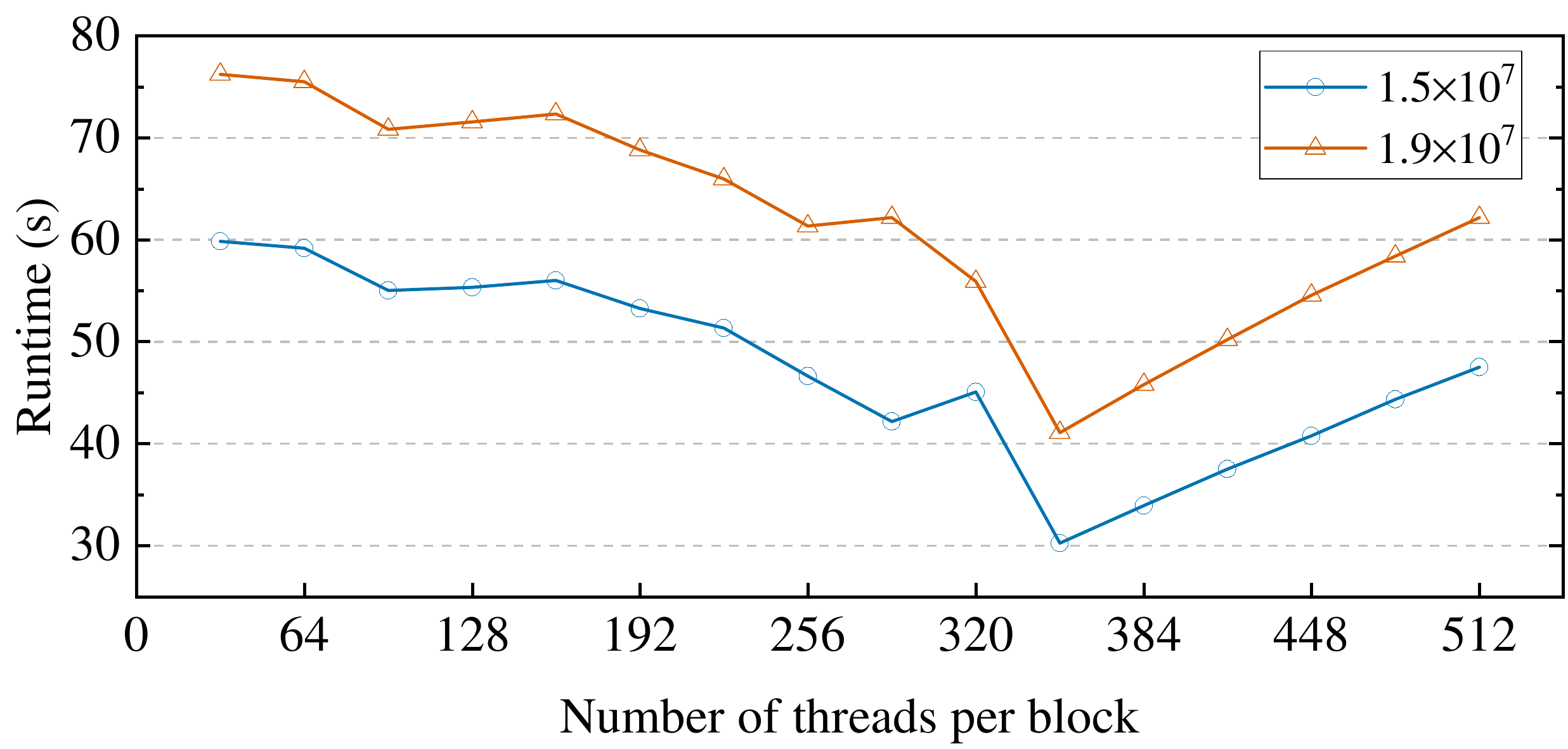}
  \caption{Performance as a function of the size of the thread blocks on NVIDIA V100 GPU. }
  \label{figure13}
\end{figure}

HEGrid vectorizes the cell update on GPU in a SIMT manner.
We evaluate the HEGrid performance by varying the size of the thread blocks on NVIDIA V100 GPU. 
The data used in this section is the simulated dataset with $1.5\times 10^7$ and $1.9\times 10^7$ data sizes, respectively.
Figure \ref{figure13} shows the HEGrid running time under different thread block sizes. 
We can observe that before reaching the optimal thread organization configuration (e.g., near 352), the performance improves as  the size of the thread block increases. This is because more threads are being scheduled to execute on the SM.
After that, when the number of threads per block is greater than 352, the running time begins a linear increase again.
The reason is that the V100 has a total number of 65,536 registers for each SM, while the HEGrid's kernel uses 88 registers as we analyzed using the nsight-compute tool \cite{wang2021grus, zhou2020tools}.  
It means that the maximum (also optimal) of block sizes that can be scheduled to execute on the SM would be 352, which could schedule two blocks to SM ( use $2 \times 352 \times 88 = 61,952$ registers, less than 65,536), that is $2\times352$ parallel threads execution on the SM.
% However, even one more warp is added, i.e., the block size is $352+32=384$, the required registers would be $2 \times 384 \times 88 \textgreater 65,536$. There would be no more blocks to be scheduled to execute on the SM and the performance degenerated.
However, even one more warp is added, i.e., the block size is $352+32=384$, there would be no more blocks (e.g., two blocks required registers would be $2 \times 384 \times 88 > 65,536$.) to be scheduled to execute on the SM, that is only 384 parallel threads execution on the SM, and the performance will degrade.
%----------------------------------------figure 13-------------------------%
%--------------------------------------------------------------------------%

\begin{figure}
  \centering
  \includegraphics[width=\linewidth]{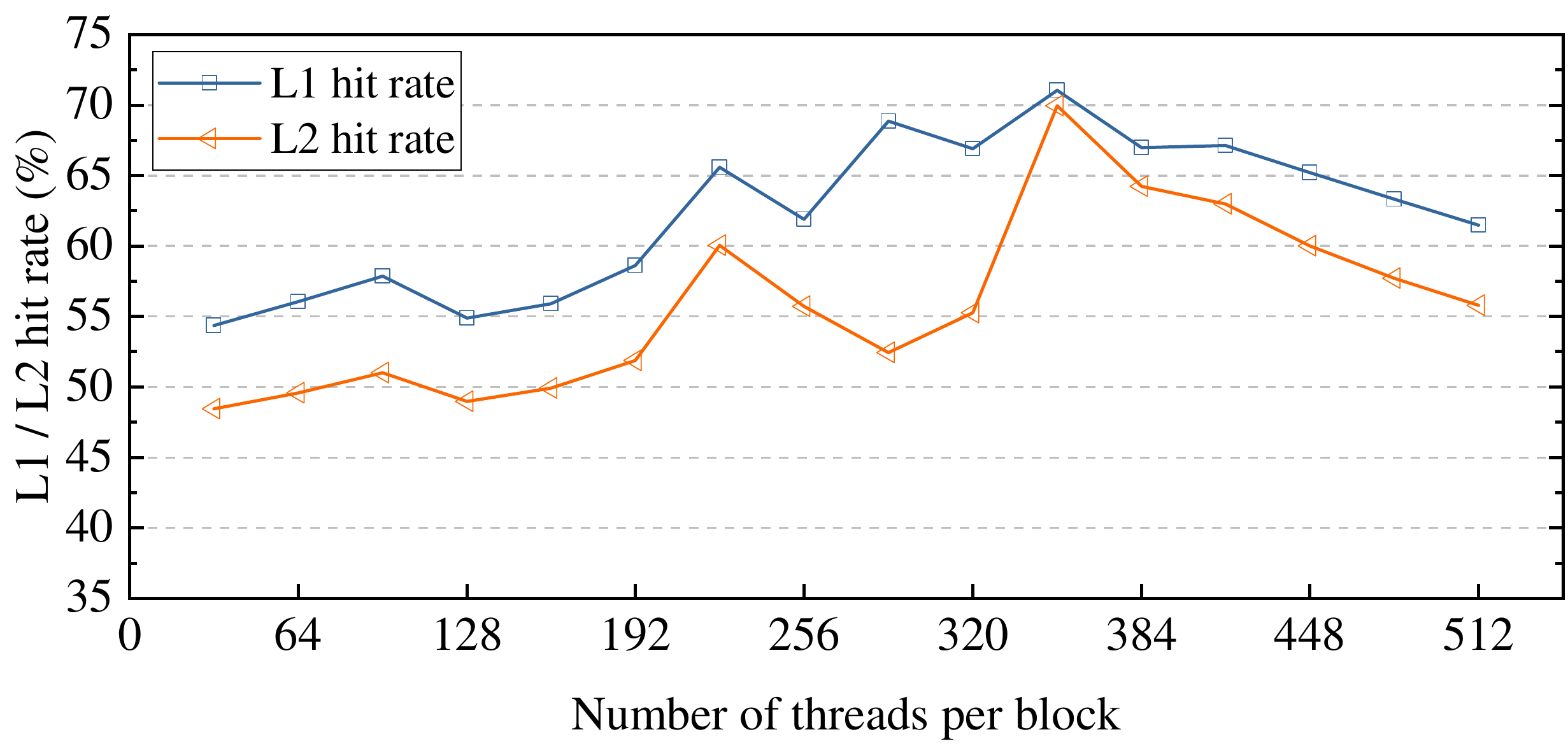}
  \caption{L1 and L2 hit rate as functions of the size of the thread blocks on NVIDIA V100 GPU.}
  \label{figure14}
\end{figure}

In addition, the GPU L1/L2 cache hit rate changes with the thread block sizes also demonstrate our thread parallelization scheme's effectiveness on GPU. 
We assign threads on the GPU in a way that takes into account the reusability of data between different threads.
As depicted in Figure \ref{figure14}, before reaching the optimal thread organization configuration (e.g., near 352), the hit rates of L1 and L2 increase with the increase of the thread block size. 
These results show that our thread organization scheme meets our expectations. 
That is, improving the GPU L1/L2 cache hit rate by organizing parallel threads through improving the inter-thread data reuse rate responsible for adjacent target cells.
%----------------------------------------figure 14-------------------------%
%--------------------------------------------------------------------------%
%----------------------------------------figure stream-------------------------%
\begin{figure*}
  \centering
  \includegraphics[width=0.95\textwidth]{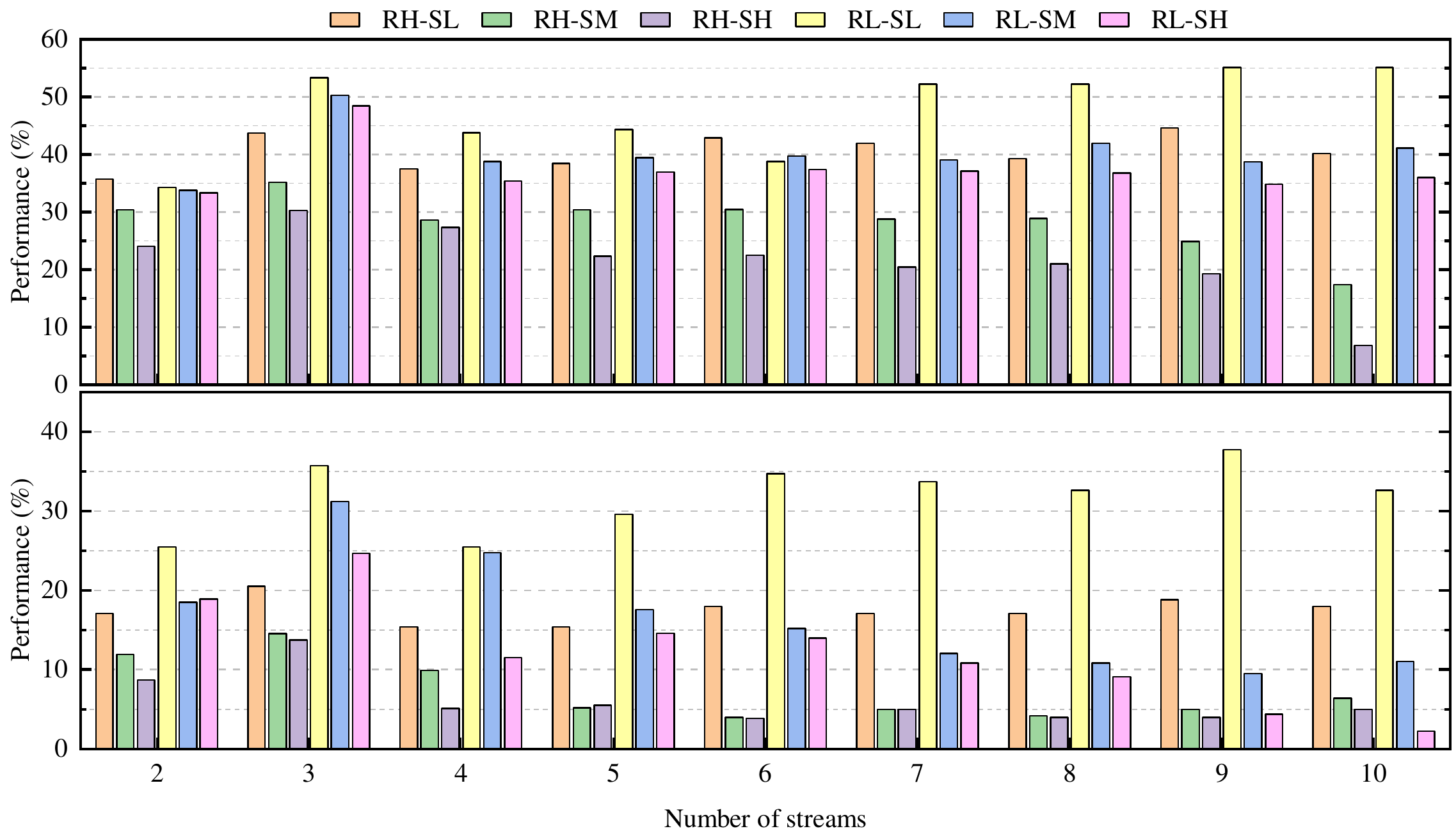}
  \caption{Performance of varied number of streams. The top and bottom represent the results of the experimental analysis for the gridding of two observation sky fields of size $5^\circ \times 5^\circ$ and $10^\circ \times 10^\circ$, respectively. The vertical axis represents the performance improvement with different streams compared to using the default stream (one stream).}
  \label{figure_stream}
\end{figure*}
%--------------------------------------------------------------------------%
\subsubsection{Streams Concurrency on GPU}
Now, we analyze the performance improvements from the multi-streams concurrency on GPU.
Here we have expanded the experimental dataset, i.e., added two sizes of observed sky fields, $5^\circ \times 5^\circ$ and $10^\circ \times 10^\circ$, and two beam widths, $180^{\prime \prime}$ and $300^{\prime \prime}$(A small beam width represents a high output resolution and more target cells on the map). 
The data size of the extended dataset range from $1.5 \times 10^5$ to $1.5 \times 10^7$.
Figure \ref{figure_stream} shows the performance improvements using different streams compared to the default stream.
Here we use "R*-S*" to represent a specific output resolution and sampling density of the experimental data, for example, RH-SH for an output resolution of $180^{\prime \prime}$ and a sampling size of $1.5 \times 10^7$, RL-SM for an output resolution of $300^{\prime \prime}$ with a sample size of $1.5 \times 10^6$, and the rest of the cases and so on.
Based on the results presented in the figure, one can make the following observations.
First, multiple streams can yield significant performance improvements compared to using the default stream, up to $55\%$ in the current experiments, which benefits from the overlapping between different streams.
Second, the performance improvement brought by the multiple concurrent streams is more pronounced in the low output resolution and small observation field cases. 
On the contrary, the performance improvement percentage tends to decrease.
This is because at low output resolutions or small observation fields, the number of cells on the output map is smaller, allowing more streams to be concurrent and a higher concurrency between different streams.
On the contrary, the computation will change to compute-intensive in the higher output resolutions or larger observation fields, and the concurrency of streams will decrease due to the resource limitation of the GPUs.
Third, the performance improvement from multiple concurrent streams is more pronounced at low sample sizes. 
On the contrary, the computation will change to memory-intensive in the larger sample size, and the concurrency of streams will decrease due to the I/O limitation.
In addition, performance improvements tend to flatten out after a threshold number of streams, which is determined by the resources of the device.

Overall, we could conclude that multi-stream concurrency could significantly improve the performance in most cases. Specifically, the optimal concurrent stream configuration needs to be tuned based on the observations, output resolution, and device resources, and dynamically adjusting the stream configuration for optimal performance is part of our future work.

\subsubsection{Thread-level Data Reuse}
\begin{figure}
  \centering
  \includegraphics[width=\linewidth]{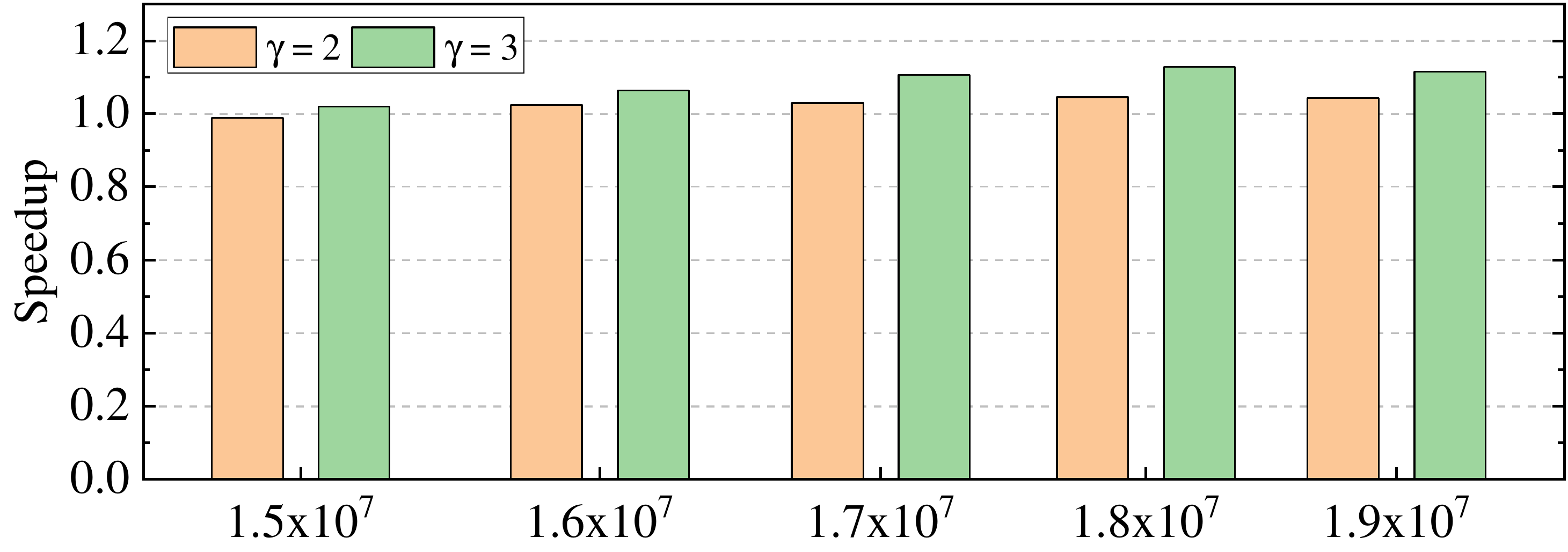}
  \caption{Performance improvement of thread-level data reuse scheme. $\gamma = 2$ and $\gamma = 3$ presents $\gamma$ adjacent cells shared same contribution region.}
  \label{figure15}
\end{figure}

We also propose a thread-level data reuse scheme targeting the high output resolution and large data size. 
Figure \ref{figure15} shows the performance improvement.
$\gamma$ is the reuse factor, representing each thread responsible for $\gamma$ adjacent grid cells. 
We can observe that, for large data sizes, thread-level data reuse scheme can achieve up to 1.2x performance speedup. 
The major reason is that it reduces the workload of the contribution points searching on the host and overhead of data loading between device memory and SM.
For example, the computation complexity changes from $\mathcal{O}(N)$ to $\mathcal{O}(N/\gamma)$, \textit{N} is the number of cells in the target map.
%----------------------------------------figure 15-------------------------%
%----------------------------------------------------------------------------%

\subsection{Performance Portability}\label{section:5.4}
To demonstrate the performance portability of HEGrid, using the simulated datasets and FAST actual observation datasets, we also evaluate HEGrid on Server\_M with AMD Instinct MI50 GPU.
The comparison of HEGrid, and Cygrid is shown in Table \ref{table4}. 
Cygrid-16 and Cygrid-32 represent Cygrid experiments using 16 and 32 CPU cores.
We can observe that running on Server\_M, HEGrid also presents promising performance, outperforming Cygrid by up to 3.8x performance speedup, which demonstrates the performance portability potential of HEGrid under different GPU architectures.
It is undeniable that HEGrid running on Server\_M exhibits a performance gap compared to running on Server\_V with V100 GPU.
The reasons behind this include.
First, the limited hardware resources of the MI50 GPU compared to the V100 GPU results in low concurrency of the HEGrid's pipelines. 
Experimental analysis reveals that for HEGrid, thread blocks can only schedule up to 128 parallel threads concurrently on the MI50 GPU SM to get "relatively better" performance.
Second, as introduced in Section \ref{section:porting}, the performance profiling tools integrated with ROCm are not yet complete, and we currently manually tune the size of thread blocks to obtain better performance. 
Therefore, besides organizing parallel threads to obtain high performance, no further performance analysis and optimizations have been done based on the architecture, which is part of our future work.
% Because MI50 GPU has limited hardware resources compared to the V100 GPU, which results in low pipeline concurrency of HEGrid, the thread block can execute only up to 128 threads concurrently on MI50 GPU, and the time-consuming increases as the channel number increases.  
% Therefore, further optimization of the pipeline concurrency combined with architecture aware is one of our future works.

%-----------------------------------Table 4----------------------------------%
% Table generated by Excel2LaTeX from sheet 'Sheet3'
\begin{table*}[width=\textwidth,cols=4,pos=h]
  \centering
  \caption{Comparison of the performance of Cygrid and HEGrid (Running on Server\_M, Running Time (s)).}
  \resizebox{\textwidth}{!}{
    \begin{tabular}{lccccccccccc}
    \toprule
    Dataset & \multicolumn{5}{c}{Simulated}         &       & \multicolumn{5}{c}{Observed (by FAST)} \\
\cmidrule{2-6}\cmidrule{8-12}    Datasize / Channel num & 1.50E+07 & 1.60E+07 & 1.70E+07 & 1.80E+07 & 1.90E+07 &       & 10    & 20    & 30    & 40    & 50 \\
    \midrule
    Cygrid-16 & 163.15 & 171.17 & 177.85 & 177.95 & 185.43 &       & 86.31 & 83.16 & 88.11 & 85.79 & 87.04 \\
    Cygrid-32 & 161.95 & 169.96 & 178.21 & 185.98 & 187.21 &       & 83.62 & 85.35 & 85.45 & 87.57 & 91.24 \\
    HEGrid & 70.75 & 75.5  & 77.42 & 80    & 85.62 &       & 21.7  & 50.88 & 78.16 & 99.56 & 125.87 \\
    \midrule
    Speedup (\textrm{HEGrid}) & \textbf{2.29} & \textbf{2.25} & \textbf{2.30} & \textbf{2.22} & \textbf{2.17} &       & \textbf{3.85} & \textbf{1.63} & \textbf{1.09} & \textbf{0.86} & \textbf{0.71} \\
    \bottomrule
    \end{tabular}}%
  \label{table4}%
\end{table*}%

\subsection{Accuracy}
Along with the performance analysis, we evaluate the accuracy of the gridding results by comparing Cygrid and HEGrid results using actual FAST observational data. 

\begin{figure*}
  \centering
  \includegraphics[width=\linewidth]{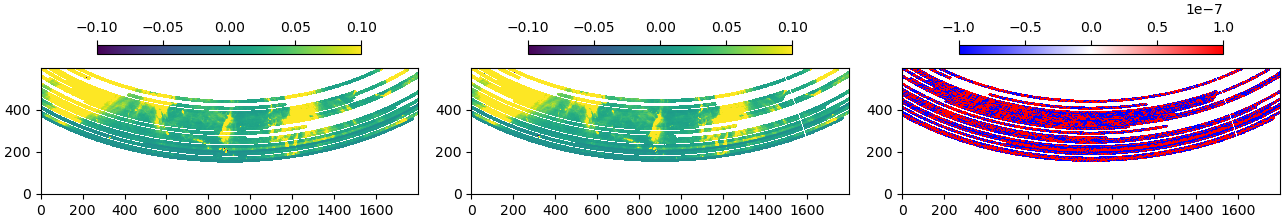}
  \includegraphics[width=\linewidth]{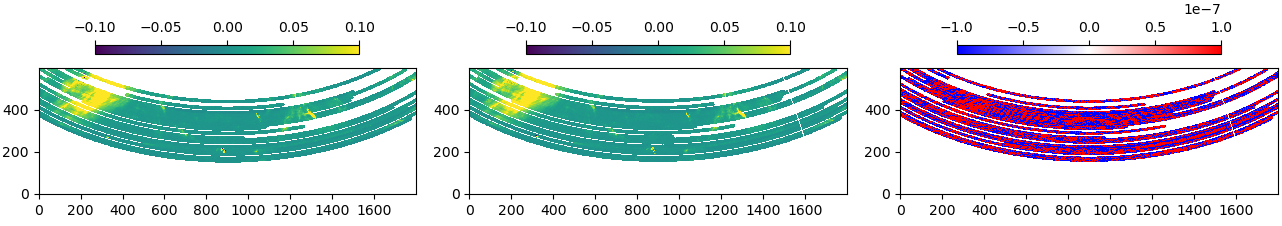}
  \caption{FAST sky images for accuracy comparison. The top and bottom represent two gridding results from two different frequency channels. The gridding results from HEGrid (left), Cygrid (middle) and their difference (right) are presented.}
  \label{figure16}
\end{figure*}

Figure \ref{figure16} shows the real sky images of one of the FAST surveys obtained from the gridding of HEGrid (left) and Cygrid (middle), respectively, and their differences (right) for comparing the accuracy of HEGrid and Cygrid.
We can observe that all-sky details can be clearly reconstructed and resolved in both cases and the difference between HEGrid and Cygrid, which is mainly caused by the different hardware architectures, is almost negligible. 
Overall, we can conclude that HEGrid retains high accuracy but better performance and is a better option for radio astronomical data gridding for large single-dish radio telescopes.  
%----------------------------------------figure 16-------------------------%
%----------------------------------------------------------------------------%

\section{Conclusions}\label{section6}
Effective and efficient data processing methods are an urgent need to fully exploit the potential of current and upcoming scientific instruments.
Gridding is the most computationally intensive step in the data reduction pipeline for data from multiple frequency channels collected by radio telescopes.
Fast and high-performance gridding frameworks of multi-channel radio astronomical data for large single-dish radio telescopes are expected to address the challenges.

In this paper, we develop a high efficient and scalable gridding framework, HEGrid, for multi-channel radio astronomical data of the large single-dish radio telescopes.
Specifically, we propose and construct the gridding pipeline in CPU-GPU heterogeneous environments and achieve multi-pipeline concurrency.
Furtherly, we propose pipeline-based co-optimization strategy to alleviate the potential negative performance impact of possible low intra- and inter-pipeline computation.
Our experiments are based on both simulated datasets and FAST's actual observed datasets. 
The results show that HEGrid shows very competitive performance compared with other state-of-the-art works.

Our future work plan for the optimization and application of HEGrid includes the following aspects.
First, for the ROCm version of HEGrid, we plan to perform detailed performance profiling and optimization based on AMD's GPU architecture, including the latest CNDA architecture, to improve performance further.
Second, we plan to achieve architecture-aware optimization, enabling HEGrid could automatically adapt to different heterogeneous architectures (CPU and GPU with different architectures) and obtain high pipeline concurrency.
Third, we plan to scale HEGrid to the cluster with multiple GPU accelerators to handle larger-scale datasets. 
Such as developing a more efficient resource scheduler in HEGrid for processing different batches of observations with varying sampling densities and sky area sizes. 
In addition, introducing HEGrid into the data reduction pipeline of FAST as a user toolkit.

\section{Declaration of competing interest}
The authors declare that they have no known competing financial interests or personal relationships that could have appeared to influence the work reported in this paper.

\section{Acknowledgments}
This work is sponsored by the Joint Research Fund in Astronomy (grant Nos.U1731125, U1731243, U1931130) under a cooperative agreement between the National Natural Science Foundation of China (NSFC) and the Chinese Academy of Sciences, NSFC grant No.11903056; as well as the National Natural Science Foundation of China (grant Nos.61972277).
\bibliographystyle{cas-model2-names}
\bibliography{references}

\begin{thebibliography}{31}
\expandafter\ifx\csname natexlab\endcsname\relax\def\natexlab#1{#1}\fi
\providecommand{\url}[1]{\texttt{#1}}
\providecommand{\href}[2]{#2}
\providecommand{\path}[1]{#1}
\providecommand{\DOIprefix}{doi:}
\providecommand{\ArXivprefix}{arXiv:}
\providecommand{\URLprefix}{URL: }
\providecommand{\Pubmedprefix}{pmid:}
\providecommand{\doi}[1]{\href{http://dx.doi.org/#1}{\path{#1}}}
\providecommand{\Pubmed}[1]{\href{pmid:#1}{\path{#1}}}
\providecommand{\bibinfo}[2]{#2}
\ifx\xfnm\relax \def\xfnm[#1]{\unskip,\space#1}\fi
%Type = Article
\bibitem[{Abdelfattah et~al.(2021)Abdelfattah, Barra, Beams, Bleile, Brown,
  Camier, Carson, Chalmers, Dobrev, Dudouit et~al.}]{abdelfattah2021gpu}
\bibinfo{author}{Abdelfattah, A.}, \bibinfo{author}{Barra, V.},
  \bibinfo{author}{Beams, N.}, \bibinfo{author}{Bleile, R.},
  \bibinfo{author}{Brown, J.}, \bibinfo{author}{Camier, J.S.},
  \bibinfo{author}{Carson, R.}, \bibinfo{author}{Chalmers, N.},
  \bibinfo{author}{Dobrev, V.}, \bibinfo{author}{Dudouit, Y.}, et~al.,
  \bibinfo{year}{2021}.
\newblock \bibinfo{title}{Gpu algorithms for efficient exascale
  discretizations}.
\newblock \bibinfo{journal}{Parallel Computing} \bibinfo{volume}{108},
  \bibinfo{pages}{102841}.
%Type = Article
\bibitem[{Bekhti et~al.(2016)Bekhti, Fl{\"o}er, Keller, Kerp, Lenz, Winkel,
  Bailin, Calabretta, Dedes, Ford et~al.}]{bekhti2016hi4pi}
\bibinfo{author}{Bekhti, N.B.}, \bibinfo{author}{Fl{\"o}er, L.},
  \bibinfo{author}{Keller, R.}, \bibinfo{author}{Kerp, J.},
  \bibinfo{author}{Lenz, D.}, \bibinfo{author}{Winkel, B.},
  \bibinfo{author}{Bailin, J.}, \bibinfo{author}{Calabretta, M.},
  \bibinfo{author}{Dedes, L.}, \bibinfo{author}{Ford, H.}, et~al.,
  \bibinfo{year}{2016}.
\newblock \bibinfo{title}{Hi4pi: a full-sky h i survey based on ebhis and
  gass}.
\newblock \bibinfo{journal}{Astronomy \& Astrophysics} \bibinfo{volume}{594},
  \bibinfo{pages}{A116}.
%Type = Article
\bibitem[{Bigot-Sazy et~al.(2015)Bigot-Sazy, Ma, Battye, Browne, Chen,
  Dickinson, Harper, Maffei, Olivari and Wilkinson}]{bigot2015hi}
\bibinfo{author}{Bigot-Sazy, M.A.}, \bibinfo{author}{Ma, Y.Z.},
  \bibinfo{author}{Battye, R.A.}, \bibinfo{author}{Browne, I.W.},
  \bibinfo{author}{Chen, T.}, \bibinfo{author}{Dickinson, C.},
  \bibinfo{author}{Harper, S.}, \bibinfo{author}{Maffei, B.},
  \bibinfo{author}{Olivari, L.C.}, \bibinfo{author}{Wilkinson, P.N.},
  \bibinfo{year}{2015}.
\newblock \bibinfo{title}{Hi intensity mapping with fast}.
\newblock \bibinfo{journal}{arXiv preprint arXiv:1511.03006} .
%Type = Article
\bibitem[{Blas et~al.(2014)Blas, Abella, Isaila, Carretero and
  Desco}]{blas2014surfing}
\bibinfo{author}{Blas, J.G.}, \bibinfo{author}{Abella, M.},
  \bibinfo{author}{Isaila, F.}, \bibinfo{author}{Carretero, J.},
  \bibinfo{author}{Desco, M.}, \bibinfo{year}{2014}.
\newblock \bibinfo{title}{Surfing the optimization space of a multiple-gpu
  parallel implementation of a x-ray tomography reconstruction algorithm}.
\newblock \bibinfo{journal}{Journal of Systems and Software}
  \bibinfo{volume}{95}, \bibinfo{pages}{166--175}.
%Type = Article
\bibitem[{C{\'a}rcamo et~al.(2018)C{\'a}rcamo, Rom{\'a}n, Casassus, Moral and
  Rannou}]{carcamo18}
\bibinfo{author}{C{\'a}rcamo, M.}, \bibinfo{author}{Rom{\'a}n, P.E.},
  \bibinfo{author}{Casassus, S.}, \bibinfo{author}{Moral, V.},
  \bibinfo{author}{Rannou, F.R.}, \bibinfo{year}{2018}.
\newblock \bibinfo{title}{Multi-gpu maximum entropy image synthesis for radio
  astronomy}.
\newblock \bibinfo{journal}{Astronomy and computing} \bibinfo{volume}{22},
  \bibinfo{pages}{16--27}.
%Type = Inproceedings
\bibitem[{Carrad et~al.(2006)Carrad, Sykes and Moorey}]{carrad2006}
\bibinfo{author}{Carrad, G.}, \bibinfo{author}{Sykes, P.},
  \bibinfo{author}{Moorey, G.}, \bibinfo{year}{2006}.
\newblock \bibinfo{title}{A cryogenically cooled seven beam 21 cm wavelength
  receiver front end for the arecibo radio telescope}, in:
  \bibinfo{booktitle}{Proc. Workshop Applications Radio Science}, pp.
  \bibinfo{pages}{15--17}.
%Type = Inproceedings
\bibitem[{Dunning et~al.(2017)Dunning, Bowen, Castillo, Chung, Doherty, George,
  Hayman, Jeganathan, Kanoniuk, Mackay et~al.}]{dunning2017}
\bibinfo{author}{Dunning, A.}, \bibinfo{author}{Bowen, M.},
  \bibinfo{author}{Castillo, S.}, \bibinfo{author}{Chung, Y.S.},
  \bibinfo{author}{Doherty, P.}, \bibinfo{author}{George, D.},
  \bibinfo{author}{Hayman, D.B.}, \bibinfo{author}{Jeganathan, K.},
  \bibinfo{author}{Kanoniuk, H.}, \bibinfo{author}{Mackay, S.}, et~al.,
  \bibinfo{year}{2017}.
\newblock \bibinfo{title}{Design and laboratory testing of the five hundred
  meter aperture spherical telescope (fast) 19 beam l-band receiver}, in:
  \bibinfo{booktitle}{2017 XXXIInd General Assembly and Scientific Symposium of
  the International Union of Radio Science (URSI GASS)},
  \bibinfo{organization}{IEEE}. pp. \bibinfo{pages}{1--4}.
%Type = Inproceedings
\bibitem[{Durrani et~al.(2021)Durrani, Chughtai, Hidayetoglu, Tahir, Dakkak,
  Rauchwerger, Zaffar and Hwu}]{durrani2021accelerating}
\bibinfo{author}{Durrani, S.}, \bibinfo{author}{Chughtai, M.S.},
  \bibinfo{author}{Hidayetoglu, M.}, \bibinfo{author}{Tahir, R.},
  \bibinfo{author}{Dakkak, A.}, \bibinfo{author}{Rauchwerger, L.},
  \bibinfo{author}{Zaffar, F.}, \bibinfo{author}{Hwu, W.m.},
  \bibinfo{year}{2021}.
\newblock \bibinfo{title}{Accelerating fourier and number theoretic transforms
  using tensor cores and warp shuffles}, in: \bibinfo{booktitle}{2021 30th
  International Conference on Parallel Architectures and Compilation Techniques
  (PACT)}, \bibinfo{organization}{IEEE}. pp. \bibinfo{pages}{345--355}.
%Type = Article
\bibitem[{Fabello et~al.(2011)Fabello, Catinella, Giovanelli, Kauffmann,
  Haynes, Heckman and Schiminovich}]{fabello2011alfalfa}
\bibinfo{author}{Fabello, S.}, \bibinfo{author}{Catinella, B.},
  \bibinfo{author}{Giovanelli, R.}, \bibinfo{author}{Kauffmann, G.},
  \bibinfo{author}{Haynes, M.P.}, \bibinfo{author}{Heckman, T.M.},
  \bibinfo{author}{Schiminovich, D.}, \bibinfo{year}{2011}.
\newblock \bibinfo{title}{Alfalfa h i data stacking--i. does the bulge quench
  ongoing star formation in early-type galaxies?}
\newblock \bibinfo{journal}{Monthly Notices of the Royal Astronomical Society}
  \bibinfo{volume}{411}, \bibinfo{pages}{993--1012}.
%Type = Article
\bibitem[{Giovanelli et~al.(2005)Giovanelli, Haynes, Kent, Perillat, Catinella,
  Hoffman, Momjian, Rosenberg, Saintonge, Spekkens
  et~al.}]{giovanelli2005arecibo}
\bibinfo{author}{Giovanelli, R.}, \bibinfo{author}{Haynes, M.P.},
  \bibinfo{author}{Kent, B.R.}, \bibinfo{author}{Perillat, P.},
  \bibinfo{author}{Catinella, B.}, \bibinfo{author}{Hoffman, G.L.},
  \bibinfo{author}{Momjian, E.}, \bibinfo{author}{Rosenberg, J.L.},
  \bibinfo{author}{Saintonge, A.}, \bibinfo{author}{Spekkens, K.}, et~al.,
  \bibinfo{year}{2005}.
\newblock \bibinfo{title}{The arecibo legacy fast alfa survey. ii. results of
  precursor observations}.
\newblock \bibinfo{journal}{The Astronomical Journal} \bibinfo{volume}{130},
  \bibinfo{pages}{2613}.
%Type = Article
\bibitem[{Gorski et~al.(2005)Gorski, Hivon, Banday, Wandelt, Hansen, Reinecke
  and Bartelmann}]{gorski05}
\bibinfo{author}{Gorski, K.M.}, \bibinfo{author}{Hivon, E.},
  \bibinfo{author}{Banday, A.J.}, \bibinfo{author}{Wandelt, B.D.},
  \bibinfo{author}{Hansen, F.K.}, \bibinfo{author}{Reinecke, M.},
  \bibinfo{author}{Bartelmann, M.}, \bibinfo{year}{2005}.
\newblock \bibinfo{title}{Healpix: A framework for high-resolution
  discretization and fast analysis of data distributed on the sphere}.
\newblock \bibinfo{journal}{The Astrophysical Journal} \bibinfo{volume}{622},
  \bibinfo{pages}{759}.
%Type = Inproceedings
\bibitem[{Griffin and Ensor(2018)}]{griffin2018end}
\bibinfo{author}{Griffin, A.}, \bibinfo{author}{Ensor, A.},
  \bibinfo{year}{2018}.
\newblock \bibinfo{title}{End-to-end modelling of the imaging pipeline in radio
  astronomy}, in: \bibinfo{booktitle}{2018 IEEE 10th Sensor Array and
  Multichannel Signal Processing Workshop (SAM)}, \bibinfo{organization}{IEEE}.
  pp. \bibinfo{pages}{480--484}.
%Type = Inproceedings
\bibitem[{Jain and Cooperman(2020)}]{jain2020crac}
\bibinfo{author}{Jain, T.}, \bibinfo{author}{Cooperman, G.},
  \bibinfo{year}{2020}.
\newblock \bibinfo{title}{Crac: checkpoint-restart architecture for cuda with
  streams and uvm}, in: \bibinfo{booktitle}{SC20: International Conference for
  High Performance Computing, Networking, Storage and Analysis},
  \bibinfo{organization}{IEEE}. pp. \bibinfo{pages}{1--15}.
%Type = Inproceedings
\bibitem[{Jung et~al.(2021)Jung, Park, Jo, Park and Lee}]{jung2021snurhac}
\bibinfo{author}{Jung, J.}, \bibinfo{author}{Park, D.}, \bibinfo{author}{Jo,
  G.}, \bibinfo{author}{Park, J.}, \bibinfo{author}{Lee, J.},
  \bibinfo{year}{2021}.
\newblock \bibinfo{title}{Snurhac: A runtime for heterogeneous accelerator
  clusters with cuda unified memory}, in: \bibinfo{booktitle}{Proceedings of
  the 30th International Symposium on High-Performance Parallel and Distributed
  Computing}, pp. \bibinfo{pages}{107--120}.
%Type = Article
\bibitem[{Leinhauser et~al.(2022)Leinhauser, Widera, Bastrakov, Debus, Bussmann
  and Chandrasekaran}]{leinhauser2022metrics}
\bibinfo{author}{Leinhauser, M.}, \bibinfo{author}{Widera, R.},
  \bibinfo{author}{Bastrakov, S.}, \bibinfo{author}{Debus, A.},
  \bibinfo{author}{Bussmann, M.}, \bibinfo{author}{Chandrasekaran, S.},
  \bibinfo{year}{2022}.
\newblock \bibinfo{title}{Metrics and design of an instruction roofline model
  for amd gpus}.
\newblock \bibinfo{journal}{ACM Transactions on Parallel Computing}
  \bibinfo{volume}{9}, \bibinfo{pages}{1--14}.
%Type = Article
\bibitem[{Li et~al.(2018)Li, Wang, Qian, Krco, Dunning, Jiang, Yue, Jin, Zhu,
  Pan et~al.}]{li18}
\bibinfo{author}{Li, D.}, \bibinfo{author}{Wang, P.}, \bibinfo{author}{Qian,
  L.}, \bibinfo{author}{Krco, M.}, \bibinfo{author}{Dunning, A.},
  \bibinfo{author}{Jiang, P.}, \bibinfo{author}{Yue, Y.}, \bibinfo{author}{Jin,
  C.}, \bibinfo{author}{Zhu, Y.}, \bibinfo{author}{Pan, Z.}, et~al.,
  \bibinfo{year}{2018}.
\newblock \bibinfo{title}{Fast in space: considerations for a multibeam,
  multipurpose survey using china's 500-m aperture spherical radio telescope
  (fast)}.
\newblock \bibinfo{journal}{IEEE Microwave Magazine} \bibinfo{volume}{19},
  \bibinfo{pages}{112--119}.
%Type = Article
\bibitem[{Merry(2016)}]{merry16}
\bibinfo{author}{Merry, B.}, \bibinfo{year}{2016}.
\newblock \bibinfo{title}{Faster gpu-based convolutional gridding via thread
  coarsening}.
\newblock \bibinfo{journal}{Astronomy and Computing} \bibinfo{volume}{16},
  \bibinfo{pages}{140--145}.
%Type = Inproceedings
\bibitem[{Otterness and Anderson(2020)}]{otterness2020amd}
\bibinfo{author}{Otterness, N.}, \bibinfo{author}{Anderson, J.H.},
  \bibinfo{year}{2020}.
\newblock \bibinfo{title}{Amd gpus as an alternative to nvidia for supporting
  real-time workloads}, in: \bibinfo{booktitle}{32nd Euromicro Conference on
  Real-Time Systems (ECRTS 2020)}, \bibinfo{organization}{Schloss
  Dagstuhl-Leibniz-Zentrum f{\"u}r Informatik}. pp.
  \bibinfo{pages}{10:1--10:23}.
%Type = Inproceedings
\bibitem[{Romein(2012)}]{romein12}
\bibinfo{author}{Romein, J.W.}, \bibinfo{year}{2012}.
\newblock \bibinfo{title}{An efficient work-distribution strategy for gridding
  radio-telescope data on gpus}, in: \bibinfo{booktitle}{Proceedings of the
  26th ACM international conference on Supercomputing}, pp.
  \bibinfo{pages}{321--330}.
%Type = Article
\bibitem[{Tr{\"a}ff et~al.(2021)Tr{\"a}ff, Hunold, Mercier and
  Holmes}]{traff2021mpi}
\bibinfo{author}{Tr{\"a}ff, J.L.}, \bibinfo{author}{Hunold, S.},
  \bibinfo{author}{Mercier, G.}, \bibinfo{author}{Holmes, D.J.},
  \bibinfo{year}{2021}.
\newblock \bibinfo{title}{Mpi collective communication through a single set of
  interfaces: A case for orthogonality}.
\newblock \bibinfo{journal}{Parallel Computing} , \bibinfo{pages}{102826}.
%Type = Inproceedings
\bibitem[{Veenboer et~al.(2017)Veenboer, Petschow and Romein}]{veenboer17}
\bibinfo{author}{Veenboer, B.}, \bibinfo{author}{Petschow, M.},
  \bibinfo{author}{Romein, J.W.}, \bibinfo{year}{2017}.
\newblock \bibinfo{title}{Image-domain gridding on graphics processors}, in:
  \bibinfo{booktitle}{2017 IEEE International Parallel and Distributed
  Processing Symposium (IPDPS)}, \bibinfo{organization}{IEEE}. pp.
  \bibinfo{pages}{545--554}.
%Type = Article
\bibitem[{Wang et~al.(2021a)Wang, Yu, Zhang, Xiao and Luo}]{wang21}
\bibinfo{author}{Wang, H.}, \bibinfo{author}{Yu, C.}, \bibinfo{author}{Zhang,
  B.}, \bibinfo{author}{Xiao, J.}, \bibinfo{author}{Luo, Q.},
  \bibinfo{year}{2021}a.
\newblock \bibinfo{title}{Hcgrid: a convolution-based gridding framework for
  radio astronomy in hybrid computing environments}.
\newblock \bibinfo{journal}{Monthly Notices of the Royal Astronomical Society}
  \bibinfo{volume}{501}, \bibinfo{pages}{2734--2744}.
%Type = Inproceedings
\bibitem[{Wang et~al.(2021b)Wang, Zhang, Li and Lin}]{wang2021exploring}
\bibinfo{author}{Wang, J.}, \bibinfo{author}{Zhang, X.}, \bibinfo{author}{Li,
  Y.}, \bibinfo{author}{Lin, Y.}, \bibinfo{year}{2021}b.
\newblock \bibinfo{title}{Exploring hw/sw co-optimizations for accelerating
  large-scale texture identification on distributed gpus}, in:
  \bibinfo{booktitle}{50th International Conference on Parallel Processing},
  pp. \bibinfo{pages}{1--10}.
%Type = Article
\bibitem[{Wang et~al.(2021c)Wang, Wang, Li, Wang, Zhu and Guo}]{wang2021grus}
\bibinfo{author}{Wang, P.}, \bibinfo{author}{Wang, J.}, \bibinfo{author}{Li,
  C.}, \bibinfo{author}{Wang, J.}, \bibinfo{author}{Zhu, H.},
  \bibinfo{author}{Guo, M.}, \bibinfo{year}{2021}c.
\newblock \bibinfo{title}{Grus: Toward unified-memory-efficient
  high-performance graph processing on gpu}.
\newblock \bibinfo{journal}{ACM Transactions on Architecture and Code
  Optimization (TACO)} \bibinfo{volume}{18}, \bibinfo{pages}{1--25}.
%Type = Inproceedings
\bibitem[{Wang et~al.(2020)Wang, Tobar, Dolensky, An, Wicenec, Wu, Dulwich,
  Podhorszki, Anantharaj, Suchyta et~al.}]{wang2020processing}
\bibinfo{author}{Wang, R.}, \bibinfo{author}{Tobar, R.},
  \bibinfo{author}{Dolensky, M.}, \bibinfo{author}{An, T.},
  \bibinfo{author}{Wicenec, A.}, \bibinfo{author}{Wu, C.},
  \bibinfo{author}{Dulwich, F.}, \bibinfo{author}{Podhorszki, N.},
  \bibinfo{author}{Anantharaj, V.}, \bibinfo{author}{Suchyta, E.}, et~al.,
  \bibinfo{year}{2020}.
\newblock \bibinfo{title}{Processing full-scale square kilometre array data on
  the summit supercomputer}, in: \bibinfo{booktitle}{SC20: International
  Conference for High Performance Computing, Networking, Storage and Analysis},
  \bibinfo{organization}{IEEE}. pp. \bibinfo{pages}{1--12}.
%Type = Article
\bibitem[{Winkel et~al.(2016)Winkel, Lenz and Fl{\"o}er}]{winkel16}
\bibinfo{author}{Winkel, B.}, \bibinfo{author}{Lenz, D.},
  \bibinfo{author}{Fl{\"o}er, L.}, \bibinfo{year}{2016}.
\newblock \bibinfo{title}{Cygrid: a fast cython-powered convolution-based
  gridding module for python}.
\newblock \bibinfo{journal}{Astronomy \& Astrophysics} \bibinfo{volume}{591},
  \bibinfo{pages}{A12}.
%Type = Article
\bibitem[{Yue et~al.(2012)Yue, Li and Nan}]{yue2012fast}
\bibinfo{author}{Yue, Y.}, \bibinfo{author}{Li, D.}, \bibinfo{author}{Nan, R.},
  \bibinfo{year}{2012}.
\newblock \bibinfo{title}{Fast low frequency pulsar survey}.
\newblock \bibinfo{journal}{Proceedings of the International Astronomical
  Union} \bibinfo{volume}{8}, \bibinfo{pages}{577--579}.
%Type = Article
\bibitem[{Zhang et~al.(2019)Zhang, Wu, Li, Kr{\v{c}}o, Staveley-Smith, Tang,
  Qian, Liu, Jin, Yue et~al.}]{zhang2019status}
\bibinfo{author}{Zhang, K.}, \bibinfo{author}{Wu, J.}, \bibinfo{author}{Li,
  D.}, \bibinfo{author}{Kr{\v{c}}o, M.}, \bibinfo{author}{Staveley-Smith, L.},
  \bibinfo{author}{Tang, N.}, \bibinfo{author}{Qian, L.}, \bibinfo{author}{Liu,
  M.}, \bibinfo{author}{Jin, C.}, \bibinfo{author}{Yue, Y.}, et~al.,
  \bibinfo{year}{2019}.
\newblock \bibinfo{title}{Status and perspectives of the crafts extra-galactic
  hi survey}.
\newblock \bibinfo{journal}{Science China Physics, Mechanics \& Astronomy}
  \bibinfo{volume}{62}, \bibinfo{pages}{1--9}.
%Type = Inproceedings
\bibitem[{Zhao et~al.(2019)Zhao, Basu, Williams, Hall and
  Johansen}]{zhao2019exploiting}
\bibinfo{author}{Zhao, T.}, \bibinfo{author}{Basu, P.},
  \bibinfo{author}{Williams, S.}, \bibinfo{author}{Hall, M.},
  \bibinfo{author}{Johansen, H.}, \bibinfo{year}{2019}.
\newblock \bibinfo{title}{Exploiting reuse and vectorization in blocked stencil
  computations on cpus and gpus}, in: \bibinfo{booktitle}{Proceedings of the
  International Conference for High Performance Computing, Networking, Storage
  and Analysis}, pp. \bibinfo{pages}{1--44}.
%Type = Inproceedings
\bibitem[{Zhou et~al.(2020)Zhou, Krentel and Mellor-Crummey}]{zhou2020tools}
\bibinfo{author}{Zhou, K.}, \bibinfo{author}{Krentel, M.W.},
  \bibinfo{author}{Mellor-Crummey, J.}, \bibinfo{year}{2020}.
\newblock \bibinfo{title}{Tools for top-down performance analysis of
  gpu-accelerated applications}, in: \bibinfo{booktitle}{Proceedings of the
  34th ACM International Conference on Supercomputing}, pp.
  \bibinfo{pages}{1--12}.
%Type = Incollection
\bibitem[{Zhu et~al.(2020)Zhu, Hou, Song, Zheng, Huang and Wu}]{zhu20}
\bibinfo{author}{Zhu, Y.}, \bibinfo{author}{Hou, J.}, \bibinfo{author}{Song,
  Y.}, \bibinfo{author}{Zheng, Y.}, \bibinfo{author}{Huang, T.},
  \bibinfo{author}{Wu, H.}, \bibinfo{year}{2020}.
\newblock \bibinfo{title}{Processing data of correlation on gpu}, in:
  \bibinfo{booktitle}{Big Data in Astronomy}. \bibinfo{publisher}{Elsevier},
  pp. \bibinfo{pages}{139--163}.

\end{thebibliography}

\end{sloppypar}
\end{document}